\documentclass[10pt,letterpaper]{article}
\usepackage[top=0.85in,left=2.75in,footskip=0.75in]{geometry}

\usepackage{amsmath,amssymb}

\usepackage{changepage}

\usepackage[utf8x]{inputenc}

\usepackage{textcomp,marvosym}

\usepackage{cite}

\usepackage{nameref,hyperref}

\usepackage[right]{lineno}

\usepackage{microtype}
\DisableLigatures[f]{encoding = *, family = * }

\usepackage[table]{xcolor}

\usepackage{array}

\usepackage{tikz}
\usepackage{algorithm}
\usepackage{algpseudocode}

\newcolumntype{+}{!{\vrule width 2pt}}

\newlength\savedwidth

\raggedright
\usepackage[aboveskip=1pt,labelfont=bf,labelsep=period,justification=raggedright,singlelinecheck=off]{caption}

\makeatletter
\renewcommand{\@biblabel}[1]{\quad#1.}
\makeatother

\usepackage{lastpage,fancyhdr,graphicx}
\usepackage{epstopdf}
\fancyheadoffset[L]{2.25in}
\fancyfootoffset[L]{2.25in}
\begin{document}
\vspace*{0.2in}

\begin{flushleft}
{\Large
\textbf\newline{A review of sequential Monte Carlo methods for real-time disease modeling} 
}
\newline
\\

Dhorasso Temfack and
Jason Wyse
\\
\bigskip

 School of Computer Science and Statistics, Trinity College Dublin, Dublin, Ireland
\\
\bigskip

%
%






\end{flushleft}
\section*{Abstract}
Sequential Monte Carlo methods are a powerful framework for approximating the posterior distribution of a state variable in a sequential manner.  They provide an attractive way of analyzing dynamic systems in real-time, taking into account the limitations of traditional approaches such as Markov Chain Monte Carlo methods, which are not well suited to data that arrives incrementally.  This paper reviews and explores the application of Sequential Monte Carlo in dynamic disease modeling, highlighting its capacity for online inference and real-time adaptation to evolving disease dynamics. The integration of kernel density approximation techniques within the stochastic Susceptible-Exposed-Infectious-Recovered (SEIR) compartment model is examined, demonstrating the algorithm's effectiveness in monitoring time-varying parameters such as the effective reproduction number. Case studies, including simulations with synthetic data and analysis of real-world COVID-19 data from Ireland,  demonstrate the practical applicability of this approach for informing timely public health interventions.



\section{Introduction}
Disease modeling plays a crucial role in society, understanding the mechanisms behind the spread of disease helps public health organizations design effective intervention strategies and support policy-making from an early stage. As the global landscape of infectious diseases continues to evolve with a growing risk of novel emerging diseases~\cite{marani2021intensity}, the need for dynamic and adaptable modeling approaches is increasingly apparent. Traditional modeling frameworks, rooted in deterministic or stochastic equations and often utilizing Markov Chain Monte Carlo (MCMC) and its variants for parameter estimation~\cite{o1999bayesian,jewell2009bayesian,baguelin2010control,kypraios2017tutorial}, have historically been employed to retrospectively analyze epidemic data and forecast the emergence of diseases. However, a major shortcoming of such modeling approaches is that they are retrospective, requiring a batch of historical data for training. It is clear that such approaches have significant shortcomings for emerging diseases which are intrinsically dynamic. Parameters of the model change during the evolution of the disease, necessitating the tracking of parameters as new data becomes available. Moreover, since various factors such as intervention strategies could affect parameters, it can be expected that the disease spread parameters change over time. Therefore, we need to develop novel approaches that can track new data as they become available. State-space models (SSM) are a general class of models that offer a versatile framework for analyzing time-series data, especially when observations can be conceptualized as noisy measurements of some unobserved latent state evolving over time. They can be used to track the time evolution of the evolving state and associated model parameters.  In the case of linear Gaussian state-space models, an exact solution to the filtering problem is provided by the Kalman filter~\cite{kalman1960new}. However, in the case of non-linear state-space models, as is the case for realistic models of most physical phenomena, analytical solutions are typically unavailable. Studies have shown that approaches based on a Kalman filter algorithm~\cite{anderson1979optimal}, are unlikely to work with high enough accuracy because of the strong non-linear setup present in such models~\cite{ chatzi2009unscented, lee2010comparison}. 

Over the past quarter of a century, there has been widespread adoption of flexible computational algorithms, collectively known as Sequential Monte Carlo (SMC) methods, especially for online state estimation. SMC methods provide a robust framework for real-time disease modeling, addressing the limitations of traditional approaches by enabling incremental data updates and dynamic parameter tracking. This paper briefly reviews joint state and parameter inference using Sequential Monte Carlo methods with a focus on epidemic modeling, providing insights into the current literature and highlighting future research directions in this field.
 
The structure of this paper is as follows: Section \ref{sec2} reviews key concepts in Bayesian particle filtering and the application of SMC methods to disease modeling. Section \ref{sec3} discusses the state-space framework for the compartmental SEIR model, focusing on sequential importance (re)sampling, and SMC filtering with unknown static parameters. Section \ref{sec4} applies the SMC methods to both simulated data and a real-world case study, specifically analyzing COVID-19 dynamics in Ireland. Finally, Section \ref{sec5} concludes the paper by summarizing the main findings and suggesting directions for future research.

\section{Sequential Monte Carlo in epidemiology}\label{sec2}
SMC methods, also known as particle filtering methods, have emerged as powerful tools for modeling dynamic systems. With an attractive and sophisticated framework, SMC enables online inference on a variety of SSMs. SMC algorithms leverage sequential importance (Re)sampling (SIS/SIR) techniques for estimating filtering distributions~\cite{Doucet2001}. Particle filters facilitate computationally efficient online inference of an evolving latent state by merging the posterior distribution at the previous time point with an incoming data batch, yielding estimates for the posterior distribution of an up-to-date latent state. The first filtering algorithm was introduced by Gordon et al.~\cite{gordon1993novel} with the name Bootstrap filter for state estimation in nonlinear, non-Gaussian state-space models and was further extended by the Auxiliary particle filter of Pitt and Shephard~\cite{pitt1999filtering}. From a computational standpoint, the SMC algorithm is faster than the standard MCMC due to its high parallelisability. Furthermore, SMC methods do not suffer from the mixing challenges that must be addressed in MCMC, especially in high-dimensional or multimodal spaces. SMC's flexibility in model specification extends its applicability to a wide range of scenarios, making it a powerful tool for dynamic systems with evolving and complex structures such as epidemiology.

The use of SMC methods for real-time disease monitoring and forecasting has a well-established history. Early attempts have demonstrated its effectiveness in calibrating nonlinear compartmental infectious disease models~\cite{ong2010realtime, dukic2012tracking, yang2014comparison, camacho2015temporal}. Findings from~\cite{safarishahrbijari2017predictive} suggest that integrating SMC algorithms with simple dynamic epidemic models can yield accurate predictions, especially in scenarios with limited parameter information but frequent batches of data over time. Advanced research has made substantial progress in utilizing SMC for disease reconstruction and forecasting, leveraging more complex data sources. Welding et al.~\cite{welding2019real} present a sequential Bayesian scheme in their work on the 2001 UK Foot-and-Mouth epidemic, exploiting the efficiency of MCMC algorithms to enable SMC to update particles and allowing real-time analysis of epidemic outbreaks using data augmentation. Additionally, Birrell et al.~\cite{birrell2020efficient} proposed an age-structured SEIR model for influenza outbreaks, showcasing the superiority of the SMC approach over MCMC for online prediction. The work of Birrell et al. is the first to address the challenge of utilizing multiple information sources to infer parameters. However, it's worth noting that the algorithms presented in these studies are not strictly sequential, as computing the posterior density up to time $t$ is required at each iteration for the MCMC move step. Given that change in the epidemic progress can be directly interpreted as a variation in the reproduction number of the infection, some papers such as~\cite{calvetti2021, won2023estimating}, have also explored the potential of SMC for tracking changes in the reproduction number of Ebola and COVID-19 outbreaks, subsequently informing public health on applicable control strategies. These studies assume that the transmission rate of the infection follows a geometric Brownian motion, and the distribution of the reproduction number is obtained through its link with the infection rate. However, their assumption of a fixed standard deviation of innovation of the transmission rate may lead to inaccurate predictions as the value is not updated when we receive new observations. A novel approach introduced by Storvik et al.~\cite{storvik2022sequential} proposes an SMC framework for inferring a time-varying reproduction number of COVID-19 in Norway. This innovative model integrates multiple information sources, including hospitalization and positive test incidences, for real-time estimation. Notably, their Auto-Regressive model stands out by adeptly capturing changes in the epidemic. It's essential to highlight that the SMC algorithm relies on a set of sufficient statistics to sequentially estimate hyperparameters in their model. Although the estimation of the parameter posterior requires an assumption that sufficient statistics-based filters are analytically tractable, this effort represents a pioneering endeavor in utilizing SMC with diverse data sources for daily estimation of the reproduction number.

\section{Compartmental model and sequential inference}\label{sec3}
Compartmental models for infectious diseases, which had their conceptual origins in the early to mid $20^{th}$ century, attempt to describe a disease's progression and spread through individuals moving through disease stages over time. The stochastic SEIR model serves as the foundation for our investigation. Consider the time interval $(t,t+\delta t]$, where $\delta t$ represents the time between observations.  If we assume that the time spent by an individual in a compartment is exponentially distributed with a rate $\lambda $, then the number of individuals leaving a compartment during the time step $\delta t$ can be seen as a Binomial variable with the probability $p=1-e^{-\lambda\delta t}$. This parameterization ensures that the probability will remain between $0$ and $1$. Hence the discrete-time  SEIR stochastic  model is given by:
\begin{align} \label{stoseir}
S_{t+ \delta t} &= S_{t} - Y_{SE}(t), \hspace{3.3cm} Y_{SE}(t) \sim\mathrm{Bin}\left(S_{t}, 1-e^{-\beta \frac{ I_{t}}{N} \delta t}\right)  \notag\\
E_{t+ \delta t} &= E_{t} + Y_{SE}(t)-Y_{EI}(t),\hspace{2cm} Y_{EI}(t) \sim\mathrm{Bin}\left(E_{t}, 1-e^{-\sigma \delta t}\right)\\
I_{t+ \delta t} &= I_{t} + Y_{EI}(t)-  Y_{IR}(t),\hspace{2.2cm} Y_{IR}(t) \sim\mathrm{Bin}\left(I_{t}, 1-e^{-\gamma  \delta t}\right) \notag\\
R_{t+ \delta t} &= R_{t} + Y_{IR}(t). \notag
\end{align}
Here, $S_t$, $E_t$, $I_t$, and $R_t$ represent the compartments of susceptible, exposed (infected but not yet infectious), infected, and recovered individuals, respectively, at time $t$. Susceptible individuals, upon contact with infected individuals, transition to the exposed class at a rate $\beta$. Exposed individuals become infectious at a rate $\sigma$, and infected individuals recover from the disease at a rate $\gamma$. The population is assumed to have a constant size $N = S_t + E_t + I_t + R_t$ at each time point.

\begin{figure}[!ht]
    \centering
        \includegraphics[scale=0.12]{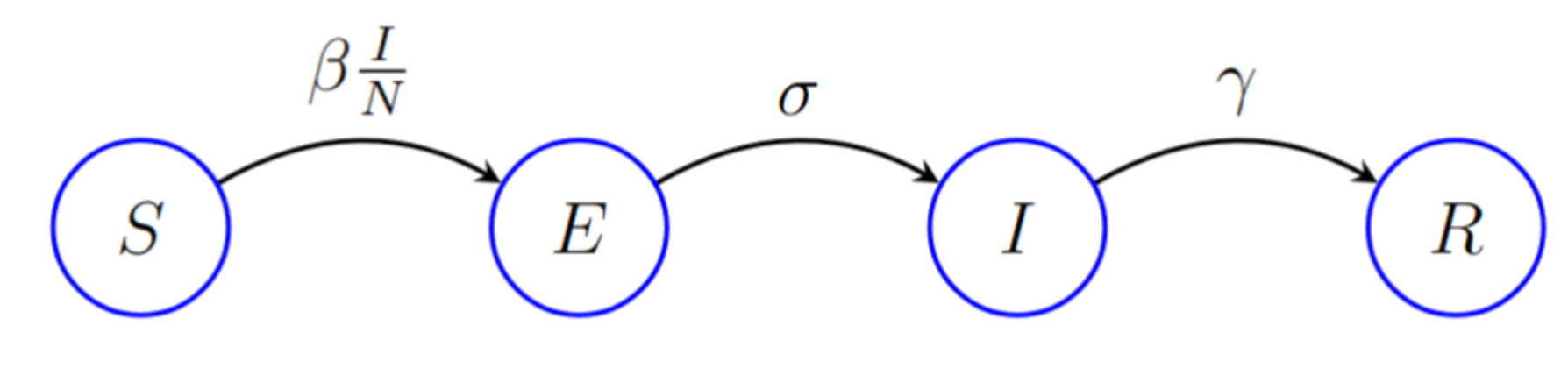}
    \caption{\footnotesize{\bf Graphical representation of a SEIR model}. Each arrow expresses the rate of transition between the compartments. $\beta$ is the disease transmission rate, $1/\sigma$ is the mean incubation period, and $1/\gamma$ is the mean infectious period.}
    \label{Fig1}
\end{figure}

\subsection{Bayesian inference in state-space models}
In the context of the SEIR model, the system can be formulated as a state-space model (SSM) or hidden Markov model (HMM), which consists of two stochastic processes: the latent state process $\{x_t\}_{t\geq0}$ and the observation process $\{y_t\}_{t\geq1}$. Here, $x_t$ represents the hidden states, specifically the compartment variables $(S_t, E_t, I_t, R_t)$, and $y_t $ represents observed data on the epidemic and. The latent states evolve according to the SEIR dynamics described in equation \eqref{stoseir}.  We can then summarise the definition of the general SSM for discrete times $t = 1, \ldots, T$ as follows:
\begin{align}
     & x_0 \sim f(x_0 |\theta),  \hspace{4cm}  \rhd  \qquad \mathrm{Initial state}
    \label{eq:obs} \\
     & x_{t}  | x_{t-1}\sim f(x_{t} | x_{t-1},\theta),  \hspace{2.5cm} \rhd  \qquad     \mathrm{State process} \label{eq:evol}     \\
    & y_{t}| x_{t} \sim g(y_{t} | x_{t},\theta).     \hspace{3.3cm}    \rhd  \qquad   \mathrm{Observation process}
    \label{eq:init}
\end{align}
Here, $f$ is the transition distribution with Markov property (see model \eqref{stoseir}). The observation $y_{t}$, is assumed to be related to a hidden state $x_{t}$ through an observation distribution $g$  and assumed independent conditional on the state $x_t$. Here, $\theta$ represents a vector of epidemiological parameters, such as  $\beta$, $\sigma$ and  $\gamma$. Throughout the study,  $\theta$ will be used interchangeably to refer to these parameters.

\begin{figure}[!ht]
  \centering
    \centering
        \includegraphics[scale=0.12]{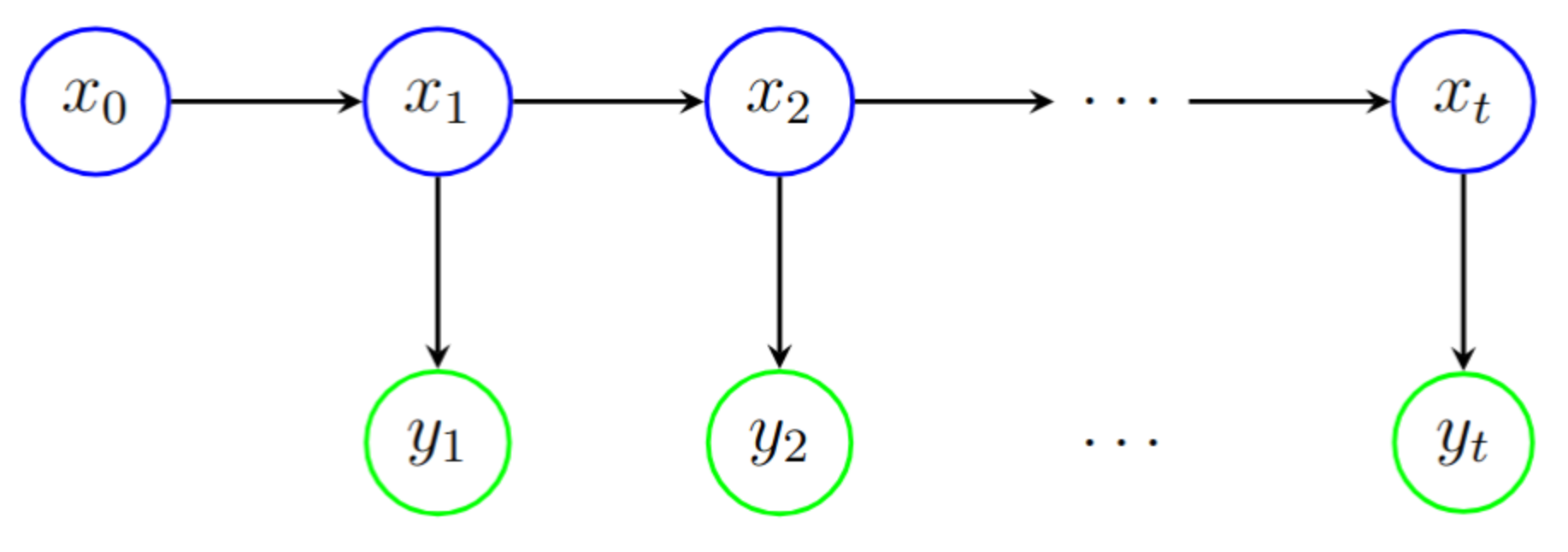}
  \caption{\footnotesize{\bf Graphical representation of an SSM}. Each arrow expresses the dependence between the variables, with the latent states in blue and the observations in green.}
\label{Fig2}
\end{figure}

 \subsubsection{Sequential Importance Sampling}
Assume that the model parameters $\theta$ are known, we will use the notation $p_{\theta}(\cdot) = p(\cdot | \theta)$ to indicate that $\theta$ conditions the probability and we denote by $x_{0:t} =  \{x_0, \ldots, x_{t} \}$ and $y_{1:t}  =  \{y_1, \ldots, y_{t} \}$. In this setting, we aim to estimate the joint smoothing distribution of the hidden states $x_{0:t}$ (i.e., the SEIR compartment counts), given all observations up to that point, $y_{1:t}$. Sequential Importance Sampling (SIS) is employed within the SMC framework to recursively estimate the filtering distribution by generating a set of weighted particles that represent the distribution. At each time step $t-1$, the joint smoothing distribution $p_{\theta}(x_{0:t-1} | y_{1:t-1})$ is approximated by a set of particles $\{x^{(i)}_{0:t-1}, W^{(i)}_{t-1}\}_{i=1}^{N_p}$, where $N_p$ denotes the number of particles and $W^{(i)}_{t-1}$ is the normalized weight associated with particle $x^{(i)}_{0:t-1}$:
\begin{equation}\label{approx}
     p_{\theta}(x_{0:t-1} | y_{1:t-1}) \approx \sum_{i=1}^{N_p} W^{(i)}_{t-1} \delta_{x^{(i)}_{0:t-1}}(x_{0:t-1}),
\end{equation}
where particles are drawn from a proposal distribution $q_{\theta}(x_{0:t-1} | y_{1:t-1})$, and the unnormalized and normalized weights are given by:
\begin{equation} \label{wt}
    w^{(i)}_{t-1} = \frac{ p_{\theta}(x_{0:t-1} | y_{1:t-1})}{q_{\theta}(x^{(i)}_{0:t-1} | y_{1:t-1})}, \quad \mathrm{ and } \quad W^{(i)}_{t-1} = \frac{w^{(i)}_{t-1}}{\sum_{i=1}^{N_p}w^{(i)}_{t-1}}.
\end{equation}
In SMC, particles $x_{0:t}^{(i)}$ are propagated using the dynamic system \eqref{stoseir}. Hence, when new data arrives, the particle filter updates the weights of the particles based on how well they are close to the observed data to approximate the joint smoothing distribution $p_{\theta}(x_{0:t}| y_{1:t})$. This is done by computing the likelihood of the observation given the state of each particle. The weight update at each time step is given by:
\begin{align} \label{wt2}
    w_t^{(i)} &\propto w_{t-1}^{(i)} \dfrac{g_{\theta}(y_t | x_t^{(i)}) f_{\theta}(x_t^{(i)} | x_{t-1}^{(i)})}{q_{\theta}(x_t^{(i)} | x_{0:t-1}^{(i)}, y_t)}, \quad  i = 1, \ldots, N_p.
\end{align}

A standard, albeit suboptimal, choice of the proposal distribution used in our applications is the prior distribution. Specifically, we set $q(x_0|\theta) = f(x_0|\theta)$ and $q(x_{t} | x_{0:t-1}, y_{t}, \theta) = f(x_{t} | x_{t-1}, \theta)$ for $t \geq 1$. Which leads to $w^{(i)}_{t}\propto w^{(i)}_{t-1} g(y_{t} | x^{(i)}_{t},\theta^{(i)}_{t})$, for $i,...,N_p$.

 \subsubsection{Particle degeneracy and resampling}
As the filtering process progresses, particles that are inconsistent with the observed data will receive lower weights and contribute minimally to the posterior mass, resulting in a small number of particles being dominant. This phenomenon is known as ``particle degeneracy''. A common approach to address this issue is to introduce a resampling step that aims to rejuvenate the diversity of particles by discarding particles with lower weights and duplicating particles with higher weights. Common resampling techniques include multinomial resampling, stratiﬁed resampling,  residual resampling and systematic resampling~\cite{gordon1993novel, kitagawa1996monte, liu1998sequential, carpenter1999improved}. Empirical studies~\cite{douc2005comparison, SHEINSON201421} have consistently shown that methods other than multinomial resampling generally offer superior performance in practical applications.
 
 An issue with resampling particles at each time point is that it may reduce the diversity amongst particles. Therefore, it is necessary to define a criterion for the resampling step when the particle weights start showing degeneracy. A common way is to estimate the degree of degeneracy by computing the effective sample size\cite{kong1994sequential}. This cannot be directly computed but can be estimated by:
 \begin{equation}
    \widehat{\mathrm{ESS}} = \dfrac{1}{\sum_{i=1}^{N_p}(W^{(i)}_{t})^2}.
\end{equation}
The effective sample size determines the number of required particles from the filtering distribution that we need to achieve the same variance as the $N_p$ particles drawn from the proposal. Then the resampling is performed only if $ \widehat{\mathrm{ESS}_{t}} < \tau_R N_p$. The threshold $ \tau_R $ is chose such that $0 \leq \tau_R \leq 1$.

\subsection{SMC  with unknown parameters
}
The framework of SMC for state inference is well-defined when all the parameters in the SSM are known. However, in many real-world scenarios, both states and parameters will be unknown and we will be interested in estimating the joint filtering density $p(x_{0:t},\theta | y_{1:t})$ at each time $t$. In this subsection, we briefly review a popular online Bayesian parameter estimation method that approximates the posterior density of the target parameter for an arbitrary dynamical system.

 Suppose that at time $t=0$, we have prior information $p(\theta)$ on our model parameters. To perform online inference in SSMs, an intuitive idea introduced by Kitagawa~\cite{kitagawa1998selforganizing},  was to augment the state by the parameters as $(x_{t}, \theta)^T$, then apply the SMC algorithm to the augmented state. Bayes's rule gives us the following marginal posterior distribution of the augmented state:
 \begin{align}
      p(x_{t},\theta | y_{1:t})
      &\propto g(y_{t}| x_{t},\theta ) p(x_{t} | \theta, y_{1:t-1})p(\theta |  y_{1:t-1}).   
 \end{align}
 However, this augmented filter is known to suffer from severe degeneracy problems. In fact, since the augmented variable  $\theta$ does not have the forgetting property \footnote{In Markovian systems, where the future behavior of the process depends only on its present state.} of the Markov process applying resampling will rapidly reduce the number of distinct $\theta$ values and lead to particle degeneracy.  To handle that, Kitagawa and Sato~\cite{kitagawa2001} assume that $\theta$ is slowly changing over time by adding artificial Gaussian noise to the parameters so that we can apply the standard SMC algorithm in the augmented state:
 \begin{equation} \label{ks}
    \theta_{t} = \theta_{t-1} + \xi_{t}, ~~\xi_{t} \sim \mathcal{N}(0, W_t).
 \end{equation} 
Where $W_t$ is some specified covariance matrix,  $\theta_{t}$\footnote{The subscript $t$ on the $\theta$ samples does not mean that $\theta$ is time-varying, it just indicates that the parameter is drawn from the posterior density at the time $t$.} and $\xi_{t}$ are assumed to be conditionally independent given the observations $y_{1:t-1}$. Although the artificial dynamic reduces the problem of particle impoverishment on the parameter vector, it has been noted that equation \eqref{ks} leads to a loss of information between time steps, resulting in greater variance in the estimated parameters than in the actual posteriors~\cite{liu_west_2001}. The accumulated error will lead to overdispersion in the approximation of the posterior distribution of parameters.

 An improved version of this method that introduces a shrinkage parameter to alleviate the overdispersion was proposed by Liu and West~\cite{liu_west_2001}.  The idea behind this approach is to use a kernel density approximation to provide  an estimate of the marginal posterior distribution parameter as a Gaussian mixture:
 \begin{equation}\label{post_theta}
  p(\theta | y_{1:t-1}) \approx \sum_{i=1}^{N_p}  W^{(i)}_{t-1} \mathcal{N}(\theta| m^{(i)}_{t-1} ,\, h^2 V_{t-1}), 
 \end{equation}
with $m^{(i)}_{t-1}= \alpha\theta^{(i)}_{t-1} + (1- \alpha)\bar{\theta}_{t-1}$, where $\bar{\theta}_{t-1}$ and $V_{t-1}$   denote the empirical mean and empirical variance-covariance matrix of the posterior samples $\left\{\theta^{(i)}_{t-1}\right\}_{i=1}^{N_p}$ at time $t-1$, respectively and given by:
 \begin{equation}
\bar{\theta}_{t-1} = \sum_{i=1}^{N_p} W^{(i)}_{t-1} \theta^{(i)}_{t-1}, \quad \mathrm{and }  \quad
V_{t-1}= \sum_{i=1}^{N_p} W^{(i)}_{t-1} \left( \theta^{(i)}_{t-1}- \bar{\theta}_{t-1} \right) \left( \theta^{(i)}_{t-1} - \bar{\theta}_{t-1} \right)^T.
 \end{equation}
The shrinkage parameter $ \alpha$ is choose such that $ \alpha = \sqrt{1-h^2}$, with  
$h^2=1-((3\delta-1)/2\delta)^2$ being the kernel smoothing parameter. The shrinkage parameter is used to shift the kernel location  $\theta^{(i)}_{t-1}$ around the overall mean of the particle set $\bar{\theta}_{t-1}$ at time $t-1$.This process will guarantee that the variance of the particles will be equal to the true posterior variance. In practice,~\cite{liu_west_2001} recommend choosing $\delta$ from the interval $(0.95, 0.99)$. 

\begin{algorithm}[!ht]
\caption{Kernel Smoothing Filter (KSF)}
Operations involving index $i$ must be performed for $i = 1,..., N_p$.

\textbf{Inputs:} Observation: $y_{1:T}$, Parameter prior: $p( \theta)$,  Proposal: $q( .|\theta)$, Discount factor: $ \delta$,  Resampling threshold.: $ \tau_R $, Number of particles: $N_p$.

\textbf{Output:}  Particles set :$\left\{(x^{(i)}_{0:t},\theta^{(i)}_{0:t}), w^{(i)}_{0:t}\right\}_{i=1}^{N_p}$.

\hrulefill 

\begin{algorithmic}[1]
    \State Sample initial particles : $x^{(i)}_0 \sim q( x_0|\theta_0)$, \quad $\theta^{(i)}_0\sim p( \theta)$
    \State Compute weights:  $w^{(i)}_0 = \dfrac{ f( x^{(i)}_0|\theta^{(i)}_0)}{q( x^{(i)}_0|\theta^{(i)}_0)}$, \quad $W^{(i)}_0 = \frac{w^{(i)}_0}{\sum_{j=1}^{N_p}w^{(j)}_0}$ 

\For{$t = 1$ to $T$}
    \If{ $\widehat{\mathrm{ESS}}< \tau_R N_p$}
    \State\label{re} Resampling: sample $\left\{ (\tilde{x}^{(i)}_{t-1},\tilde{\theta}^{(i)}_{t-1})\right\}_{i=1}^{N_p}$ from $\left\{ (x^{(i)}_{t-1},\theta^{(i)}_{t-1})\right\}_{i=1}^{N_p}$ with the probability $\left\{W^{(i)}_{t-1}\right\}_{i=1}^{N_p}$
    \EndIf 
    \State Propagate parameter:  $\theta^{(i)}_{t}\sim  \mathcal{N}( \alpha\tilde{\theta}^{(i)}_{t-1} + (1- \alpha)\bar{\theta}_{t-1} ,\, h^2 V_{t-1})$
    \State Propagate state: $x^{(i)}_{t} \sim q(x^{(i)}_{t} | \tilde{x}^{(i)}_{0:t-1}, y_{t}, \theta^{(i)}_{t})$ 
    \State Update weights:  $w^{(i)}_{t}\propto w^{(i)}_{t-1}\dfrac{ g(y_{t} | x^{(i)}_{t},\theta^{(i)}_{t}) f(x^{(i)}_{t} | \tilde{x}^{(i)}_{t-1},\theta^{(i)}_{t}) }{q(x^{(i)}_{t} | \tilde{x}^{(i)}_{0:t-1}, y_{t},\theta^{(i)}_{t})}$, ~ $W^{(i)}_{t} =\frac{ w^{(i)}_{t}}{\sum_{j=1}^{N_p}w^{(j)}_{t}}$
\EndFor

\end{algorithmic}
\label{Alg1KSF}
\end{algorithm}

 \section{Experimental and real-world data analysis}\label{sec4}

\subsection{Experimental setup}
We first apply the SMC algorithm to two synthetic datasets generated using the stochastic SEIR model outlined at the beginning of Section 3. Experiment 1 considers a situation where all the parameters of the model are constant and Experiment 2 examines a time-varying transmission rate.

\subsubsection{Experiment 1}
Synthetic data were generated using the stochastic model  \eqref{stoseir} for a total population $N=6000$. For this case, we suppose all the parameters in the SEIR stochastic model are static and set as $\beta =0.45$, $\sigma = 1/3$, and $\gamma = 1/5$, with an initial condition $(S_0, E_0, I_0, R_0 )=(N-1,0,1,0)$. The simulation period was set $T=120$ days with a time step $\delta t=1$ day.  As suggested in~\cite{SHEINSON201421}, we used stratified resampling, with  $10,000$ particles and the discount factor $\delta=0.99$. Here our latent state is the vector $x_{t}=(S_{t}, E_{t}, I_{t}, R_{t} )^T$, with the vector of unknown parameters  $\theta=(\beta, \sigma, \gamma )^T$ to be estimated. The observation of daily new infections is modeled using the Poisson distribution, which offers an intuitive interpretation for generating daily count events on a given day, denoted as $y_{t}|x_{t}\sim\mathrm{Poisson}(Y_{EI}(t))$. The value of the initial state vector in our algorithm is set as $(S_0, E_0, I_0, R_0 )=(N-I_0,0, I_0,0)$, with discrete uniform $I_0\sim\ \mathcal{U}(\{0,\dots,3\})$. 
We assume independent priors $\beta\sim \mathcal{U}([0.1, 0.6])$, $\sigma\sim \mathcal{U}([1/14, 1/2])$ and $\gamma\sim\mathcal{U}([1/15, 1/2])$. A log transform is applied to $\beta, \sigma$ and $\gamma$ in order to work with the KSF mixture. This accommodates the strictly positive support of these parameters.

\begin{figure}[!ht]
		\begin{center}
                \includegraphics[scale=0.55]{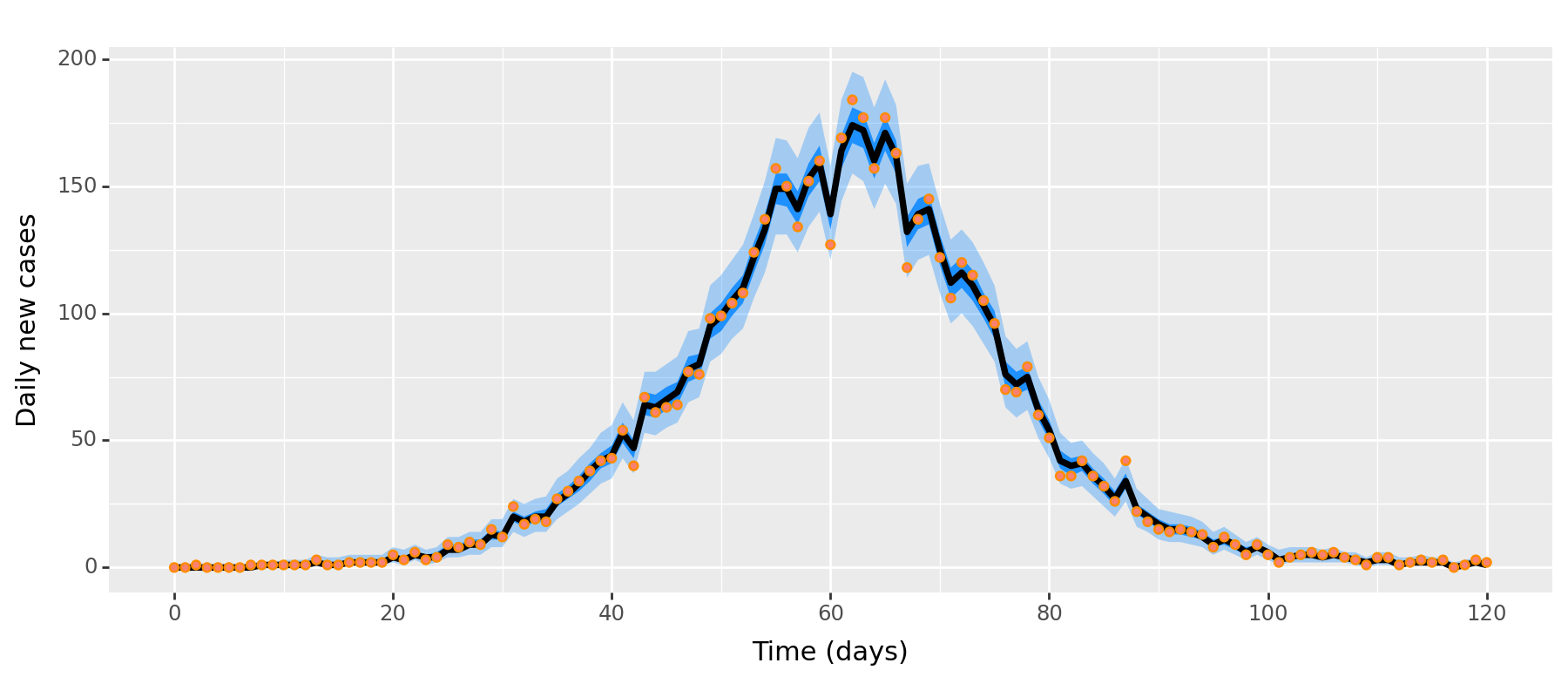}
                
                \includegraphics[scale=0.43]{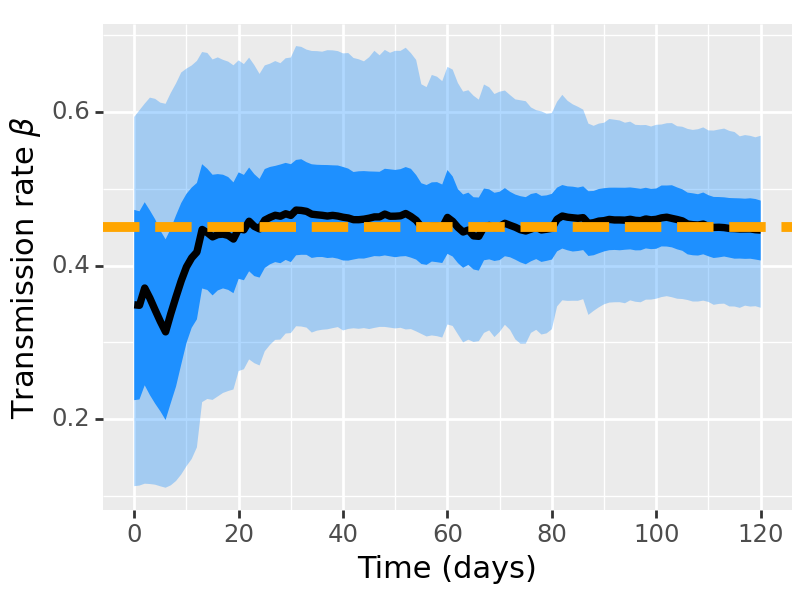}
                \includegraphics[scale=0.43]{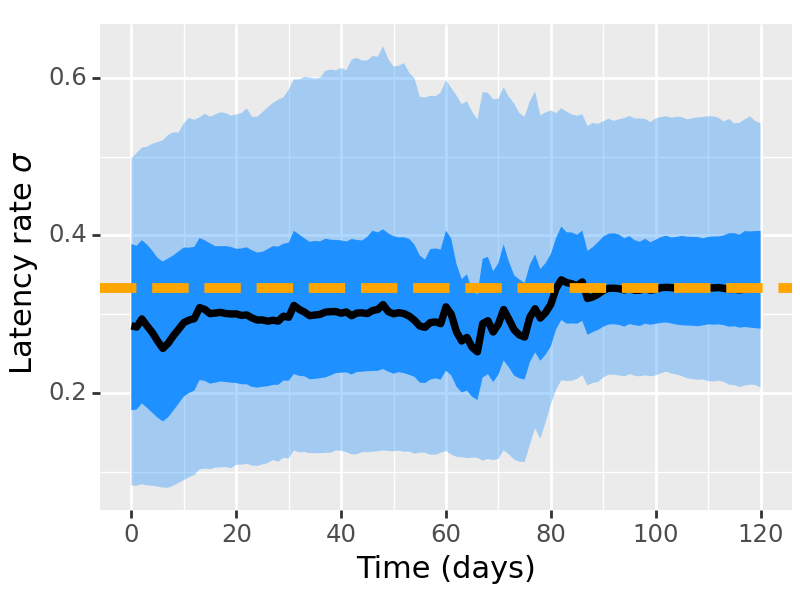}
                \includegraphics[scale=0.43]{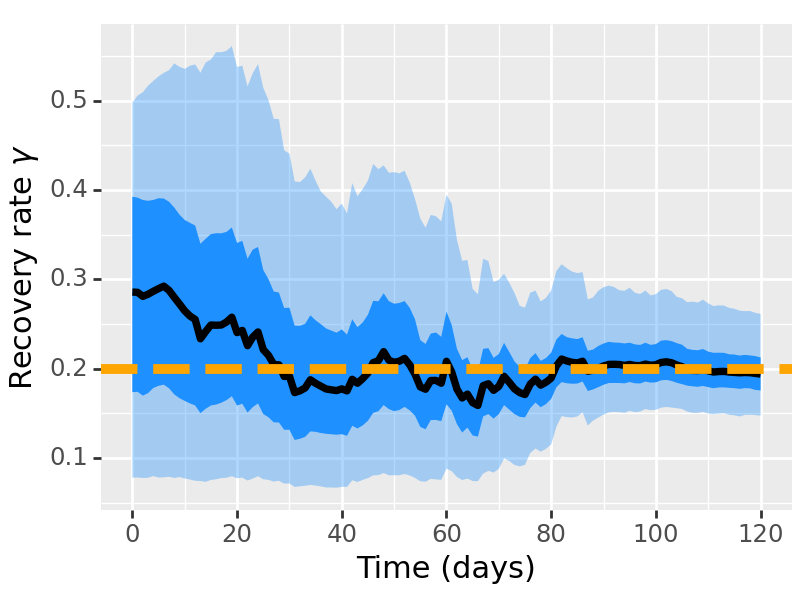}

                \includegraphics[scale=0.43]{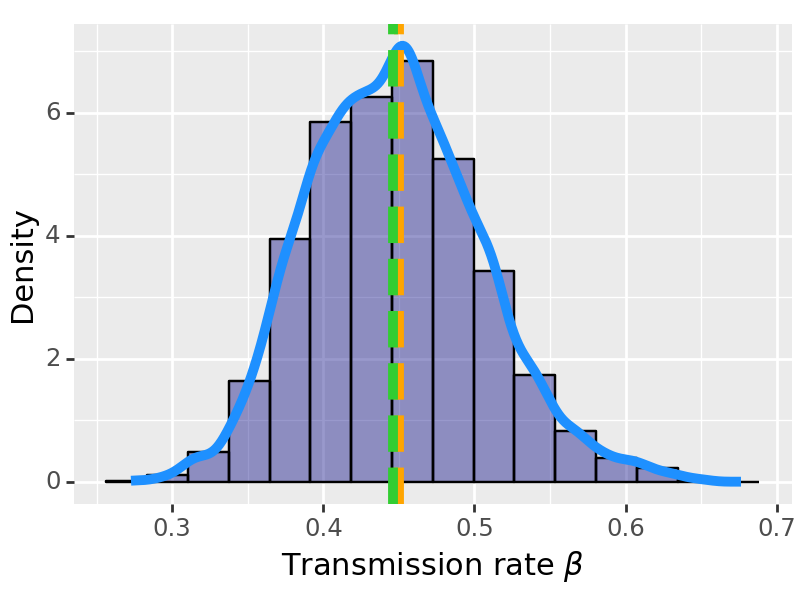}
                \includegraphics[scale=0.43]{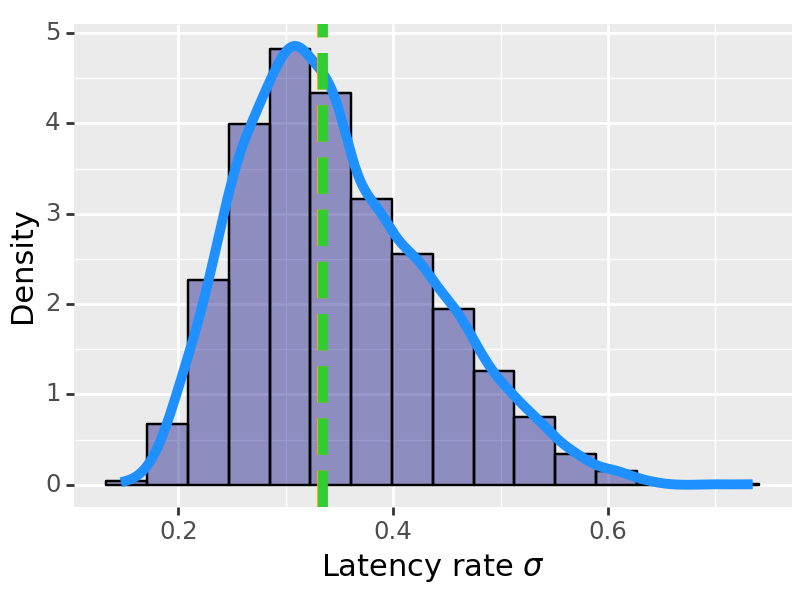}
                \includegraphics[scale=0.43]{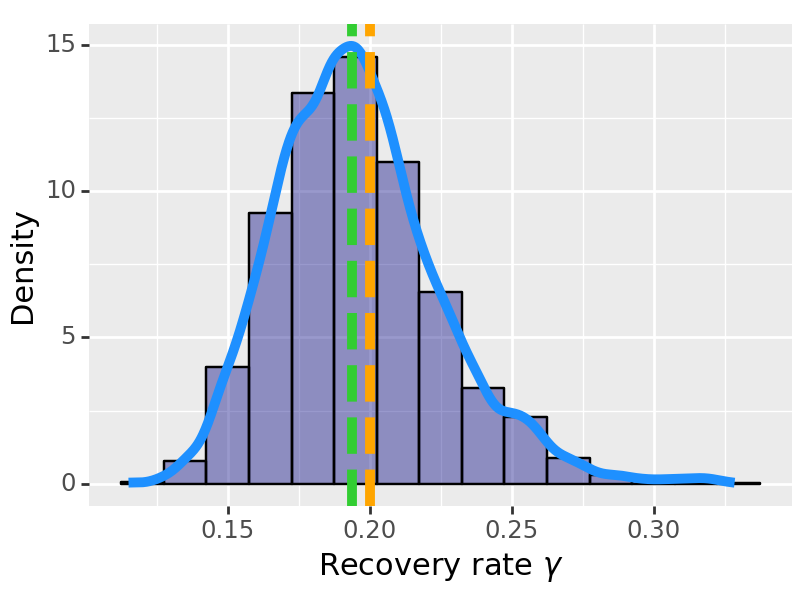}
 	\end{center}
\caption{\footnotesize {\bf State and parameters inference for Experiment 1}. Top row: SMC estimate of the daily new infected and the true observations are represented by orange circle dots. Middle row: Filtering estimate of the parameters $\beta$, $\sigma$, and $\gamma$ over time.
Bottom row: posterior distributions for the parameters $\beta$, $\sigma$, and $\gamma$ at the last time step, the green dashed line indicates the posterior mean. The true parameter values by horizontal and vertical orange dashed lines in rows 2 and 3.  The solid black line, the dark blue and light blue area correspond to the posterior median, the 50\% and 95\% credible intervals, respectively.}
\label{Fig3}
\end{figure}

Fig~\ref{Fig3} displays the Bayesian SMC inference of daily new cases of infection together with the static parameters. We can observe that the predicted new cases are in good agreement with the data. The results also indicate that all true parameter values fall within the 50\% credible interval, closely aligning with the posterior median/mean as we evolve over time. Initially, during the epidemic's onset, the data lacks sufficient information within the time interval $[0,25]$, resulting in widespread parameter uncertainty. However, as we get closer to the peak of the epidemic, the inference algorithm captures the dynamics of the epidemic, notably reducing parameter uncertainty.

\subsubsection{Experiment 2}
To mimic a more realistic scenario, we introduce variability in the transmission rate. Hence the synthetic data were generated assuming that the transmission rate is a sinusoidal function of time ($\beta(t)=0.5(1+0.75\sin (0.2t))$,  with  $\sigma=1/4$ and $\gamma=1/5$. Since the transmission rate is no longer constant, we model $\beta_{t}$ using a geometric Brownian motion:
\begin{equation} \label{rw}
    \beta_{t}=e^{\log (\beta_{t-1})+\varepsilon_{t}}, \quad \varepsilon_{t}\sim\mathcal{N}(0,\nu_{\beta}^2).
\end{equation}
Here, $\nu_{\beta}$ is the parameter controlling the volatility and will be estimated. Hence  our latent state is the vector $x_{t}=(S_{t}, E_{t} , I_{t}, R_{t}, \beta_{t})^T$, with the vector of unknown parameters  $\theta=(\sigma , \gamma , \nu_{\beta})^T$. This example aims to evaluate how well the KSF can capture the state and parameters when there is high variability in disease spread dynamics. Parameters used to run the KSF algorithm were kept at the same value as in Experiment 1 with the initial transmission rate $\beta_0\sim\mathcal{U}(0.3, 0.6)$. The inverse gamma prior for the squared volatility, $\nu_{\beta}^2$, is regularly used in Bayesian dynamic systems~\cite{arias2021bayesian}. Specifically, we set $\nu_{\beta}^2 \sim \mathcal{IG}(26, 0.03)$ to allow for rapid changes in the transmission rate, ensuring that the model can adapt to sudden shifts in disease dynamics. This choice also aligns with values observed in previous studies~\cite{won2023estimating, funk2018realtime}.

\begin{figure}[!ht]
		\begin{center}
                \includegraphics[scale=0.55]{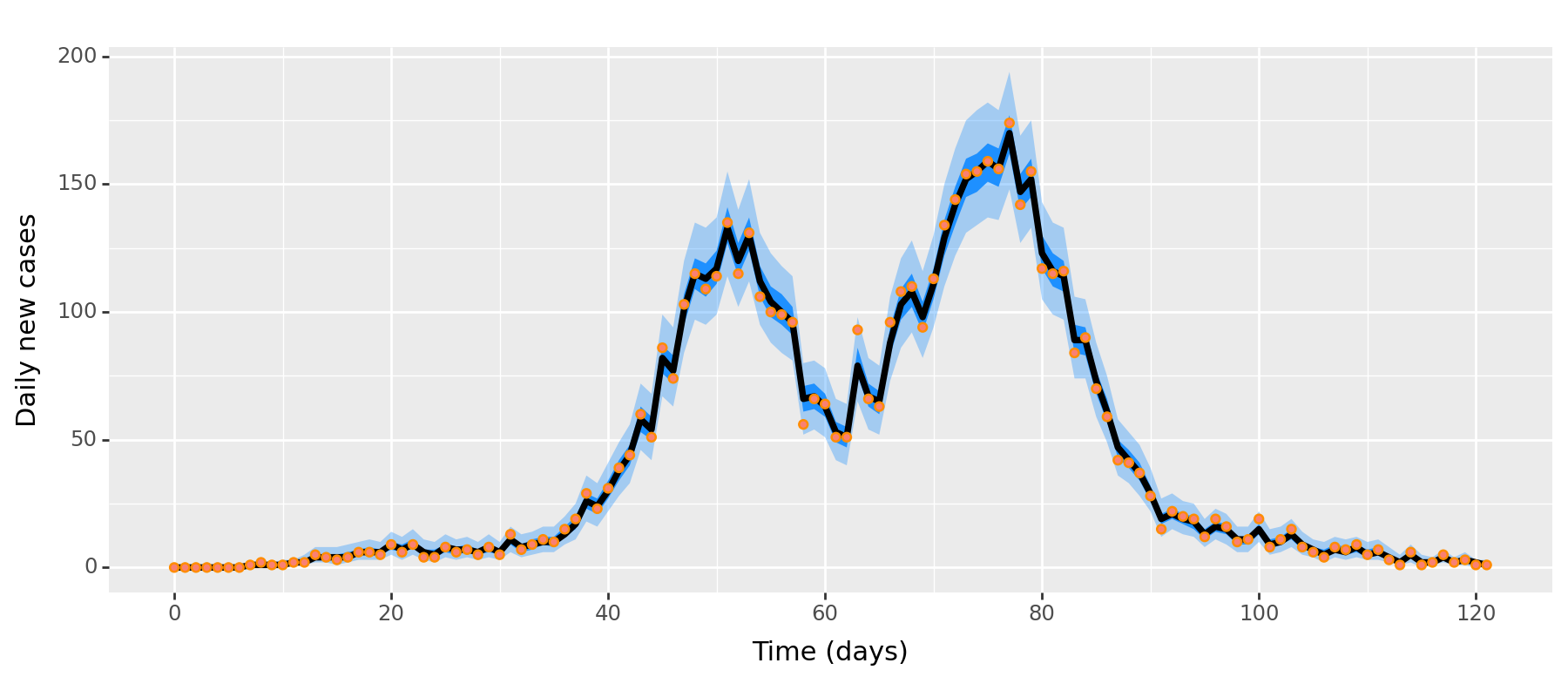}
                
                \includegraphics[scale=0.43]{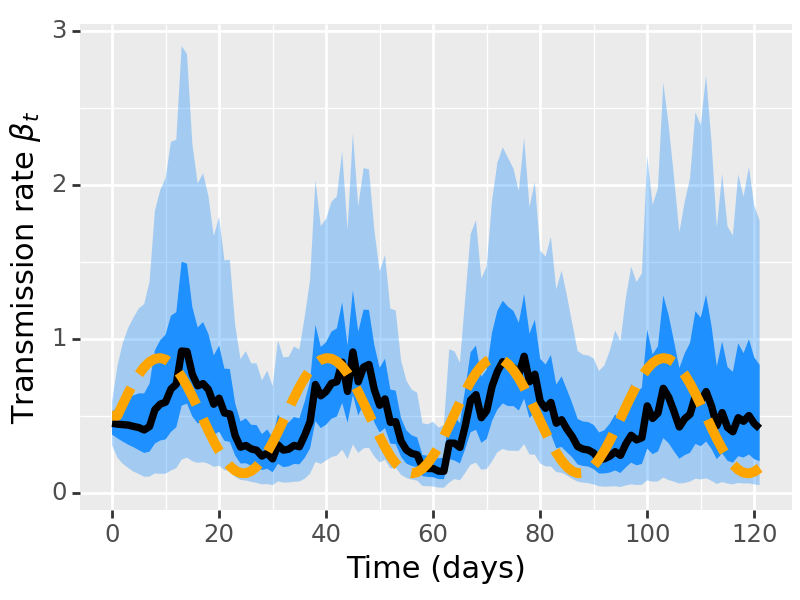}
                \includegraphics[scale=0.43]{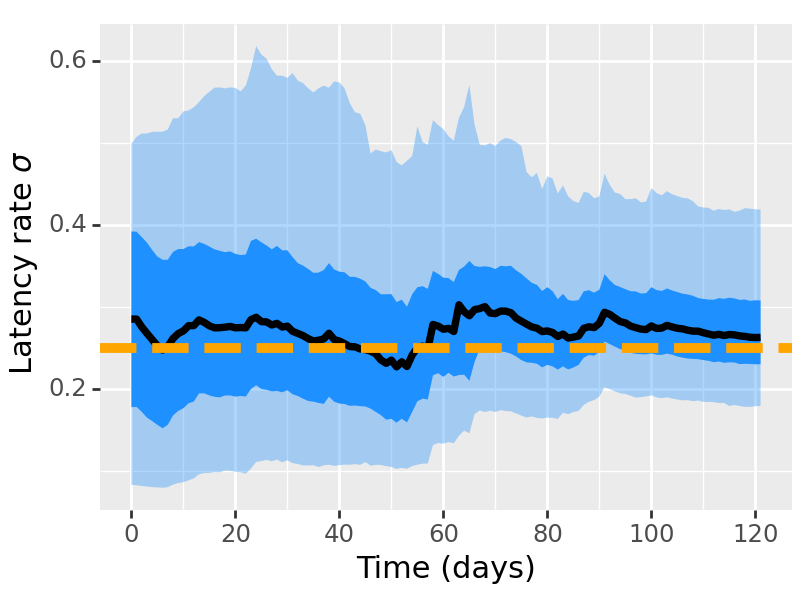}
                \includegraphics[scale=0.43]{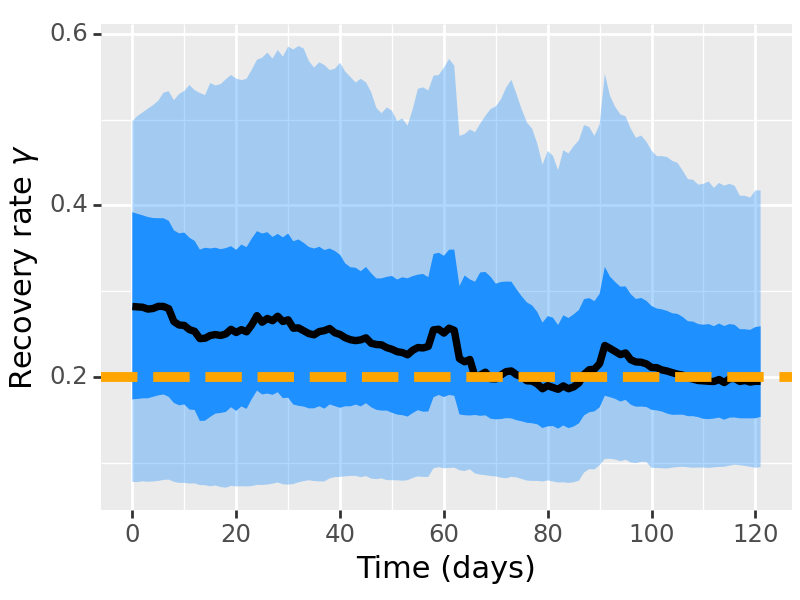}

                \includegraphics[scale=0.43]{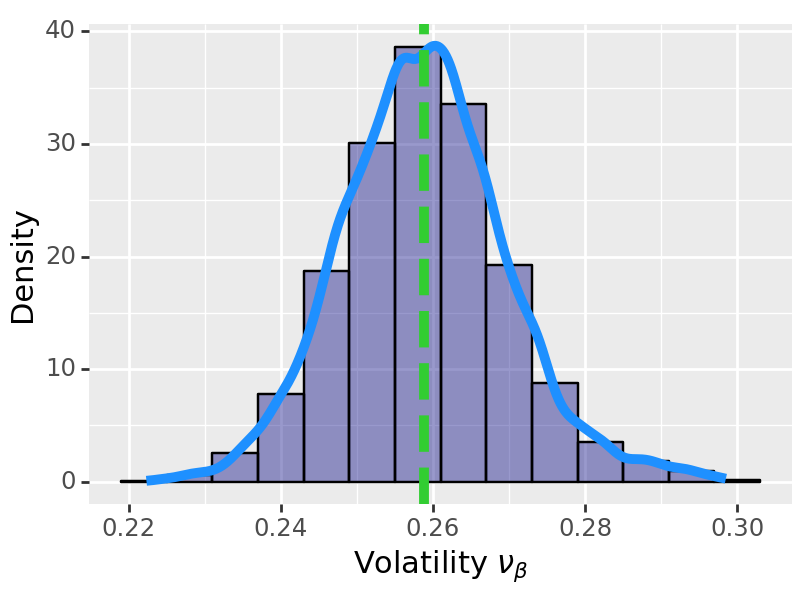}
                \includegraphics[scale=0.43]{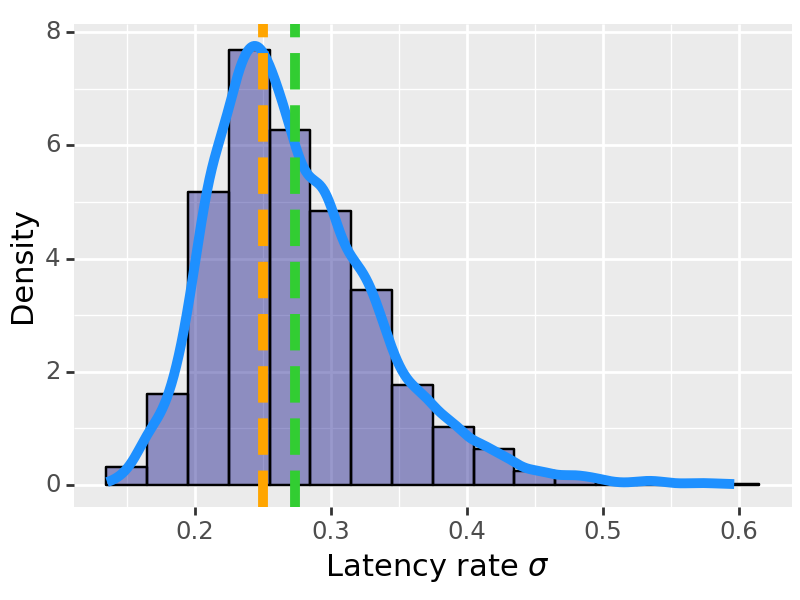}
                \includegraphics[scale=0.43]{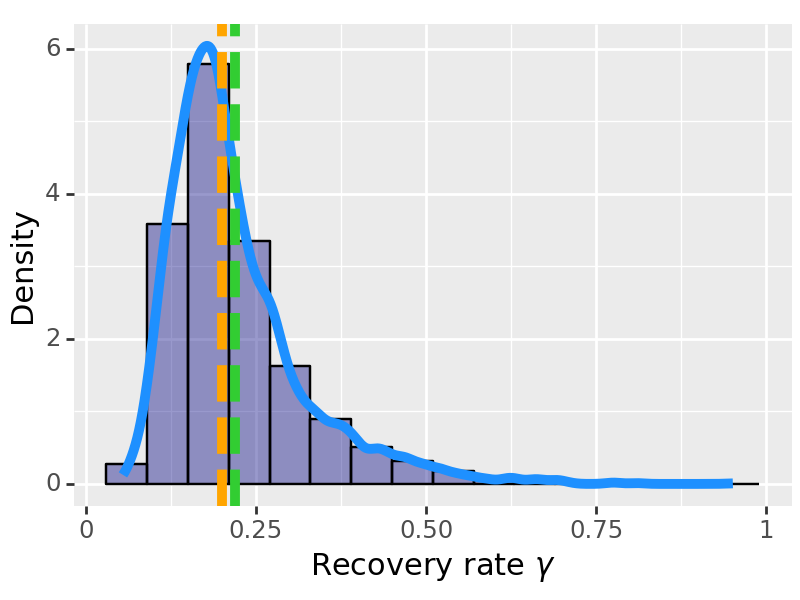}
 	\end{center}
	\caption{\footnotesize {\bf State and parameters inference for Experiment 2}. Top row: SMC estimate of the daily new infected and the true observations are represented by orange circle dots. Middle row: Filtering estimate of the parameters $\beta$, $\sigma$, and $\gamma$ over time. Bottom row: posterior distributions for the parameters $\nu_{\beta}$, $\sigma$, and $\gamma$ at the last time step, the green dashed line indicates the posterior mean. The true parameter values by horizontal and vertical orange dashed lines in rows 2 and 3.  The solid black line, the dark blue and light blue area correspond to the posterior median, the 50\% and 95\% credible intervals, respectively.}\label{Fig4}
\end{figure}

In Fig~\ref{Fig4}, the model's ability to track the various waves of an outbreak and the overall trend of a time-varying transmission rate is demonstrated, albeit with a short delay. This delay is attributed to the latency inherent in the daily reported case data (approximately proportional to the mean incubation period $1/\sigma$).  As in the previous experiment, the true values of the static parameters fall within the 50\% credible interval, indicating robust parameter estimation.

\subsection{Real case study: Analysis of COVID-19 in Ireland}

Since the onset of the COVID-19 pandemic in December 2019 in Wuhan, China, the World Health Organization (WHO) has reported over 772 million confirmed cases globally and more than 69 million deaths~\cite{who2020covid}. The rapid transmission and widespread nature of the virus led the WHO to declare it a pandemic. In~\cite{cazelles2021dynamics}, particle MCMC has been employed to reconstruct the COVID-19 epidemic's trajectory in Ireland. However, it is worth noting that while this method efficiently tracks observed data, it remains an offline approach and may not be suitable for real-time decision-making when rapid responses are required.

\subsubsection{Stochastic model}

We present a fully discrete stochastic version of the continuous and deterministic extended-SEIR model introduced by the  Irish Epidemiological Modelling Advisory Group (IEMAG) ~\cite{IEMAG2020} to study the spread of COVID-19. The model encompasses various compartments: $S_{t}$ denotes susceptible individuals, $E_{t}$ represents infected individuals not yet infectious, $I^{ps}_{t}$ and $I^{as}_{t}$ represent the pre-symptomatic and asymptomatic infected individuals, respectively. Individuals self-isolating without testing are labeled $I^{si}_{t}$. To account for delays in reporting new cases, symptomatic individuals are divided into two compartments: Symptomatic individuals awaiting test results are denoted as $I^{st}_{t}$, while those symptomatic and isolating after receiving positive test results are $I^{pi}_{t}$. Symptomatic individuals not tested or isolated are represented by $I^{sn}_{t}$, and finally, $R_{t}$ is the count of individuals recovering or succumbing to the virus. The model is described by the following discrete equations:
\begin{align} \label{covm}
    S_{t+\delta t} &= S_{t} - Y_1(t), \notag \\
    E_{t+\delta t} &= E_{t} + Y_1(t) - Y_2(t), \notag \\
    I^{ps}_{t+\delta t} &= I^{ps}_{t} + Y_3(t) - Y_4(t), \notag \\
    I^{as}_{t+\delta t} &= I^{as}_{t} + Y_2(t) - Y_3(t) - Y_7(t), \notag \\
    I^{si}_{t+\delta t} &= I^{si}_{t} + Y_5(t) - Y_8(t), \\
    I^{st}_{t+\delta t} &= I^{st}_{t} + Y_6(t) - Y_9(t), \notag \\
    I^{pi}_{t+\delta t} &= I^{pi}_{t} + Y_9(t) - Y_{10}(t), \notag \\
    I^{sn}_{t+\delta t} &= I^{sn}_{t} + Y_4(t) - Y_5(t) - Y_6(t) - Y_{11}(t), \notag \\
    R_{t+\delta t} &= R_{t} + Y_7(t) + Y_8(t) + Y_{10}(t) + Y_{11}(t), \notag 
\end{align}
where $Y_i(t), ~i=1,...,11$ are random variables with binomial distributions defined by:
\begin{adjustwidth}{-0.7in}{0in} 
\begin{align} \label{covb}
&Y_1(t) \sim \mathrm{Bin}\left(S_{t}, 1 - \mathrm{exp}\left(-\beta_{t} (I^{ps}_{t} + r_{as} I^{as}_{t} + r_{si} I^{si}_{t} + I^{st}_{t} + r_{pi}  I^{pi}_{t} + I^{sn})\cdot\delta t/N\right)  \right), \quad\notag \\
&Y_2(t) \sim \mathrm{Bin}\left(E_{t}, 1 - e^{-\frac{1}{\tau_l} \delta t}\right), \quad Y_3(t) \sim \mathrm{Bin}\left(Y_2(t), 1 -f_{as}\right), \quad
Y_4(t) \sim \mathrm{Bin}\left(I^{ps}_{t}, 1 - e^{-\frac{1}{\tau_c-\tau_l} \delta t}\right), \quad\notag \\
&Y_5(t) \sim \mathrm{Bin}\left(Y_4(t), f_{si}\right), \quad Y_6(t) \sim \mathrm{Bin}\left(Y_4(t), f_{st} \right), \quad
Y_7(t) \sim \mathrm{Bin}\left(I^{as}_{t}, 1 - e^{-\frac{1}{\tau_d} \delta t}\right), \quad\notag \\
&Y_8(t) \sim \mathrm{Bin}\left(I^{si}, 1 - e^{-\frac{1}{\tau_d-\tau_c+\tau_l} \delta t}\right), \quad Y_9(t) \sim \mathrm{Bin}\left(I^{st}_{t}, 1 - e^{-\frac{1}{\tau_r} \delta t}\right), \quad\notag \\
&Y_{10}(t) \sim \mathrm{Bin}\left( I^{pi}_{t}, 1 - e^{-\frac{1}{\tau_d-\tau_c+\tau_l-\tau_r} \delta t}\right), \quad Y_{11}(t) \sim \mathrm{Bin}\left(I^{sn}_{t}, 1 - e^{-\frac{1}{\tau_d-\tau_c+\tau_l} \delta t}\right).
\end{align} 
\end{adjustwidth}
The model assumes homogeneous population mixing and a constant size: 
\begin{equation}
 N = S_{t} + E_{t} + I^{as}_{t} + I^{ps}_{t} + I^{si}_{t} + I^{st}_{t} + I^{pi}_{t} + I^{sn}_{t} + R_{t}.   
\end{equation}

Let $D_t$ be the number of deceased individuals, then it can be expressed as a fraction of the removed individuals as follows:
\begin{equation}
D_{t+\delta t}= D_t + Y_{12}(t),~~Y_{12}(t)\sim\mathrm{Bin}\left(R_{t}, d_t\right)
\end{equation}
 where $d_t$ is a time-varying death probability following the relation:
\begin{equation}
\mathrm{logit}(d_t)=\mathrm{logit}(d_{t-1}) + \varepsilon_{t}, \quad \varepsilon_{t} \sim \mathcal{N}(0, \nu_{d}^2).
\end{equation}
Here  $\mathrm{logit}(d_t)=\log\left(\frac{d_t}{1-d_t}\right)$, is the logit-transform which ensures that the death probability is constrained to $(0,1)$. $\varepsilon_{t}$ represents the random fluctuation in the death probability at time $t$, following a normal distribution with mean zero and variance $\nu_{d}^2$. This formulation effectively captures the dynamic nature of the death probability over time, enabling us to account for improvements in clinical outcomes as time progresses and we set the initial death probability $d_0\sim\mathcal{U}(0.001, 0.002)$.

The models presented in Eq~\eqref{covm}-\eqref{covb}, as well as the IEMAG model~\cite{IEMAG2020}, do not explicitly incorporate vaccination. This is because they are primarily designed to focus on the spread of the virus and the impact of various non-pharmaceutical interventions during a period when vaccines are either unavailable or not yet widely deployed.  However, the effect of the effect of vaccination can easily be introduced by adding a compartment for vaccinated individuals and adjusting the transmission dynamics to reflect partial or complete immunity. Allowing the transmission rate to be time-varying seems reasonable since, throughout the outbreak, the intensity of disease transmission will tend to vary due to individuals' behavioral changes or government lockdown measures. Hence, we consider a geometric random walk (RW) model for the transmission rate:
\begin{equation} \label{rwm} 
    \beta_{t} = e^{\log (\beta_{t-1}) + w_{t}}, \quad w_{t} \sim \mathcal{N}(0, \nu_{\beta}^2),
\end{equation}
where $\nu_{\beta}$ is the parameter controlling the innovation in the transmission rate and we set $\beta_0\sim\mathcal{U}(0.7, 0.8)$, consistent with the study of COVID-19 in Ireland in~\cite{cazelles2021dynamics}.

\subsubsection{Data and model fitting}
The data source used to make inferences was restricted to daily confirmed cases and cumulative death counts extracted from the updated database of the Health Protection Surveillance Centre (HPSC) (\href{https://COVID19ireland-geohive.hub.arcgis.com/}{https://COVID19ireland-geohive.hub.arcgis.com/}). To integrate these observations into our model, we define two variables: $C^I_{t}$ representing the cumulative number of infectious cases and $C^D_{t}$ representing cumulative deaths. These are given by the following equations:
\begin{equation}
    C^I_{t+1} = C^I_{t} + Y_9(t),~ \mathrm{ and }  C^D_{t+1} = C^D_{t} + D_t
\end{equation}
Hence, $\mu_{t} = C^I_{t+1} - C^I_{t}$ is the daily expected number of cases. Since the reported cases are subject to observational errors, and we observe that the sample variance is larger than the sample mean in the data, the daily observed cases ($Z^I_t$) are modeled using the normal approximation of a negative binomial distribution of mean $\mu_{t}$ and variance $\mu_{t}(1+\phi \mu_{t})$, where $\phi$ denotes the overdispersion parameter. To prevent singularity when, $\mu_{t}=0$, we followed the approach outlined in~\cite{funk2018realtime}, rounding variances smaller than $1$ up to $1$. For actual cumulative death count ($Z^D_t)$, we assumed that  it follows a Poisson distribution with mean $ C^D_t$:
\begin{align}
Z^I_t\sim\mathrm{Normal}(\mu_{t},\mu_{t}(1+\phi \mu_{t})), \mathrm{ and }
     Z^D_t\sim\mathrm{Poisson}(C^D_t).
\end{align}
We estimate the spread of COVID-19 in Ireland from February 29th, 2020, to March 31st, 2021, using the daily new infected count data of 396 days and cumulative death count data. The Irish population in 2020 was estimated to be $N= 4,965,439$.  The KSF uses $N_p =20,000$ particles with stratified resampling at each time step. The discount factor is chosen as $\delta=0.99$. Table~\ref{Table1} gives the description of the parameters used in the model \eqref{covb}.

\begin{table}[!ht]
\begin{adjustwidth}{-0.4in}{0in} 
\caption{\footnotesize {\bf Description of the parameters of the model \eqref{covm}-\eqref{rwm}}.
The value of the parameters $\tau_c$, $\tau_l$, $\tau_d$  and $\tau_r$ are taken at their average.} \label{Table1}
\begin{tabular}{c| l l l l}
\hline
\textbf{Parameter} & \textbf{Description} & \textbf{Value}&\textbf{Reference}\\
\hline\hline
$r_{as}$ & Factor reduction of transmission from $I^{as}$ &0.55&\cite{jaouimaa2021age}\\
$r_{si}$ &  Factor reduction of transmission from $I^{si}$&0.05 &\cite{jaouimaa2021age}\\
$r_{pi}$ &  Factor reduction of transmission from $I^{pi}$  &0.05&\cite{jaouimaa2021age}\\
$\tau_l$ & Average latent period  &4.9  days &~\cite{IEMAG2020}\\
$\tau_c$ & Average incubation period &5.85 days&~\cite{IEMAG2020}\\
$\tau_d$ & Average infectious period &7  days &~\cite{IEMAG2020}\\
$\tau_r$ & Average period from first symptom to test result&3.52 days&~\cite{IEMAG2020}\\
$f_{as}$ & fraction of infected asymptomatic&0.2&\cite{jaouimaa2021age}\\
$f_{si}$ & fraction  of symptomatic self-isolated&0.1 &\cite{jaouimaa2021age}\\
$f_{st}$ & fraction  of symptomatic already tested &0.8 &\cite{jaouimaa2021age}\\
\hline
\end{tabular}
\end{adjustwidth}
\end{table}
 The hyperparameters in the particle filter were randomly generated from a prior distribution. We selected the inverse gamma distribution for $ \nu_{d}^2 $ and $ \nu_{\beta}^2$. Specifically, we set $\nu_{d}^2 \sim \mathcal{IG}(80, 0.02)$ and $\nu_{\beta}^2 \sim \mathcal{IG}(70, 0.02)$, aligning with prior range of value in found in~\cite{arias2021bayesian}.  These choices enable our model to take into account the significant variability observed in the empirical data. For the overdispersion parameter, we adopted $\phi \sim \mathcal{IG}(35, 0.2)$.This is consistent with the results of previous research by Funk et al.~\cite{funk2018realtime} on Ebola, which highlighted the fact that a large overdispersion value effectively captures trajectory variability.  Regarding initial states, we set: $ S_{0} =N- E_{0}-I^{ps}_{0}$, $E_{0}=1$, $I^{ps}_{0}\sim\mathcal{U}(\{5,\dots,100\})$ and $ I^{as}_{0}=I^{si}_{0} =I^{st}_{0} =I^{pi}_{0} = I^{sn}_{0}= R_{0}=0$.
\begin{figure}[!ht]
	\begin{center}
            \includegraphics[scale=0.47]{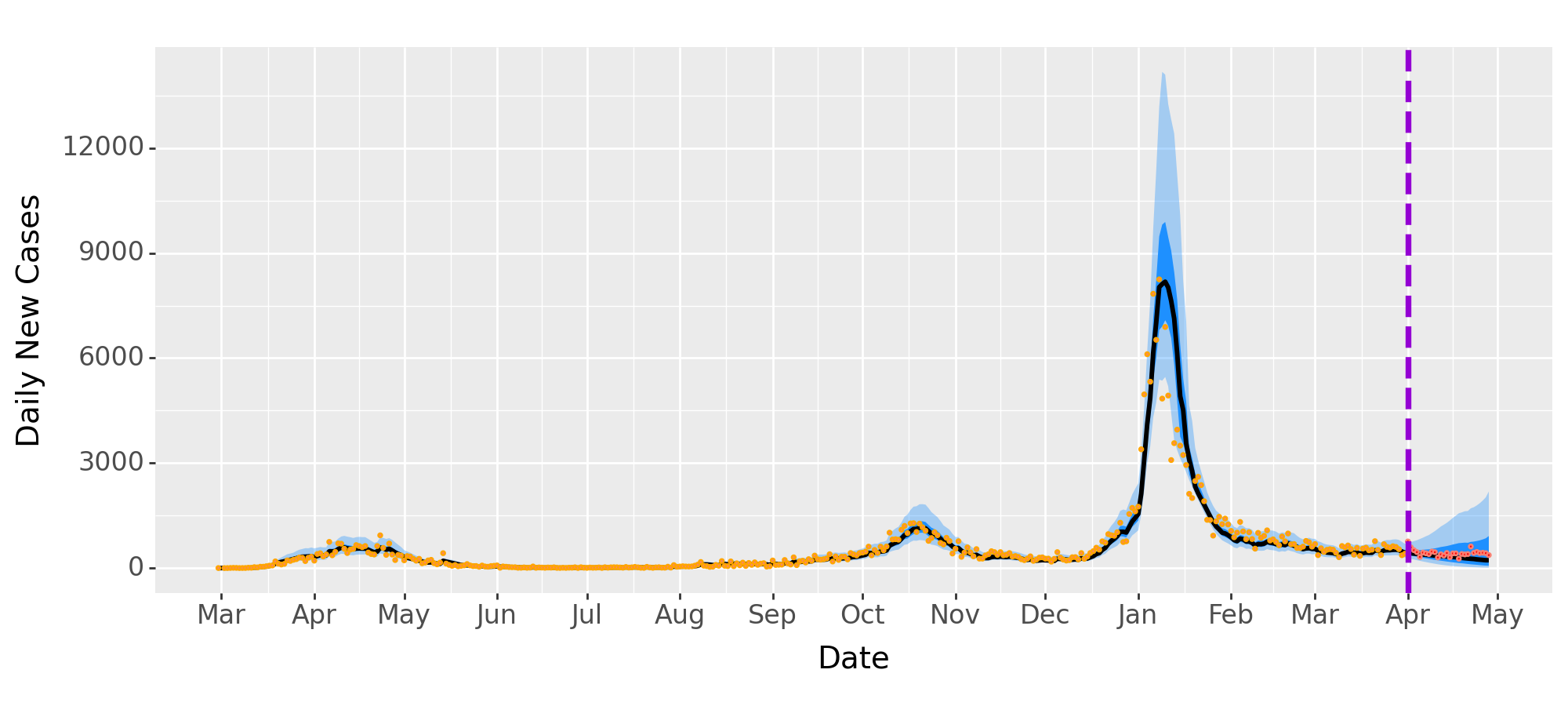}
            \includegraphics[scale=0.47]{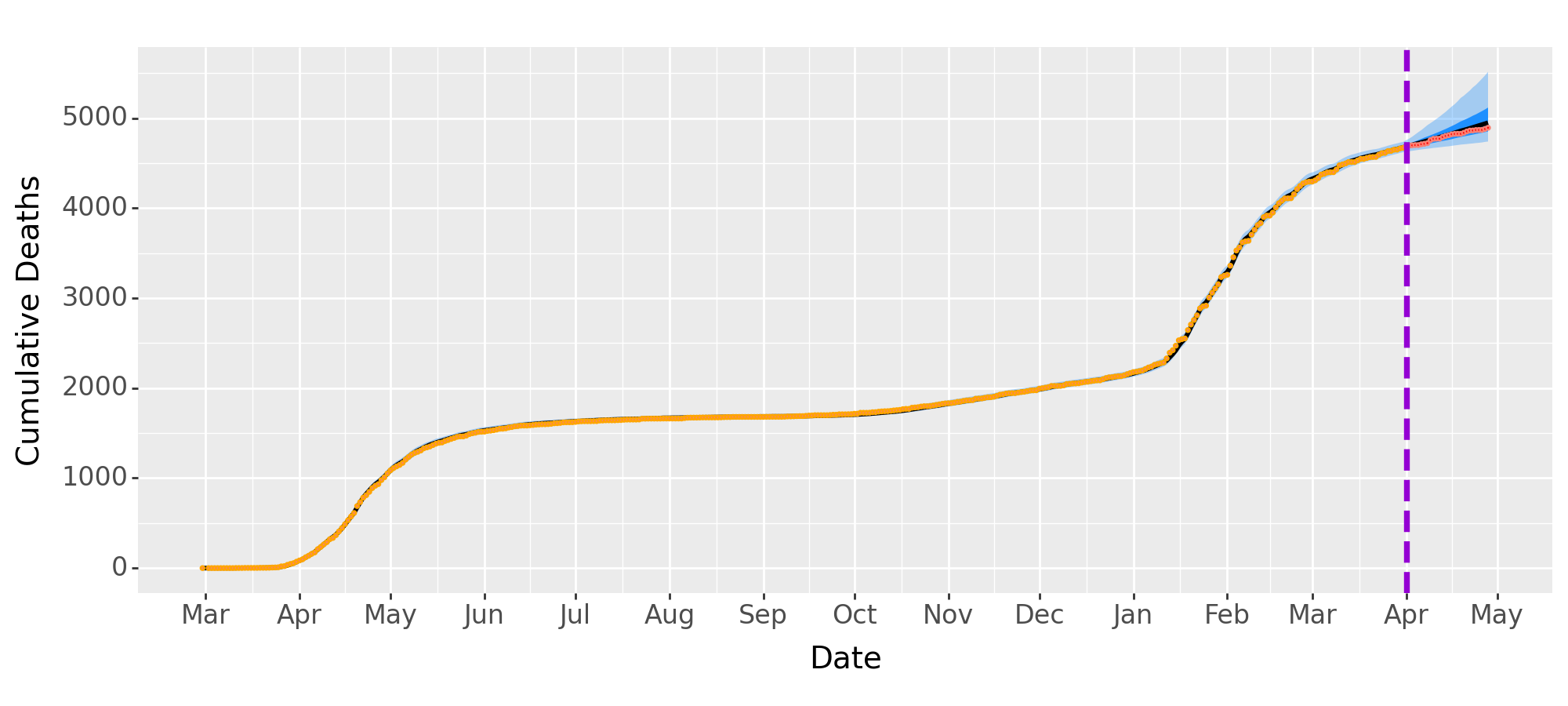}
            \includegraphics[scale=0.47]{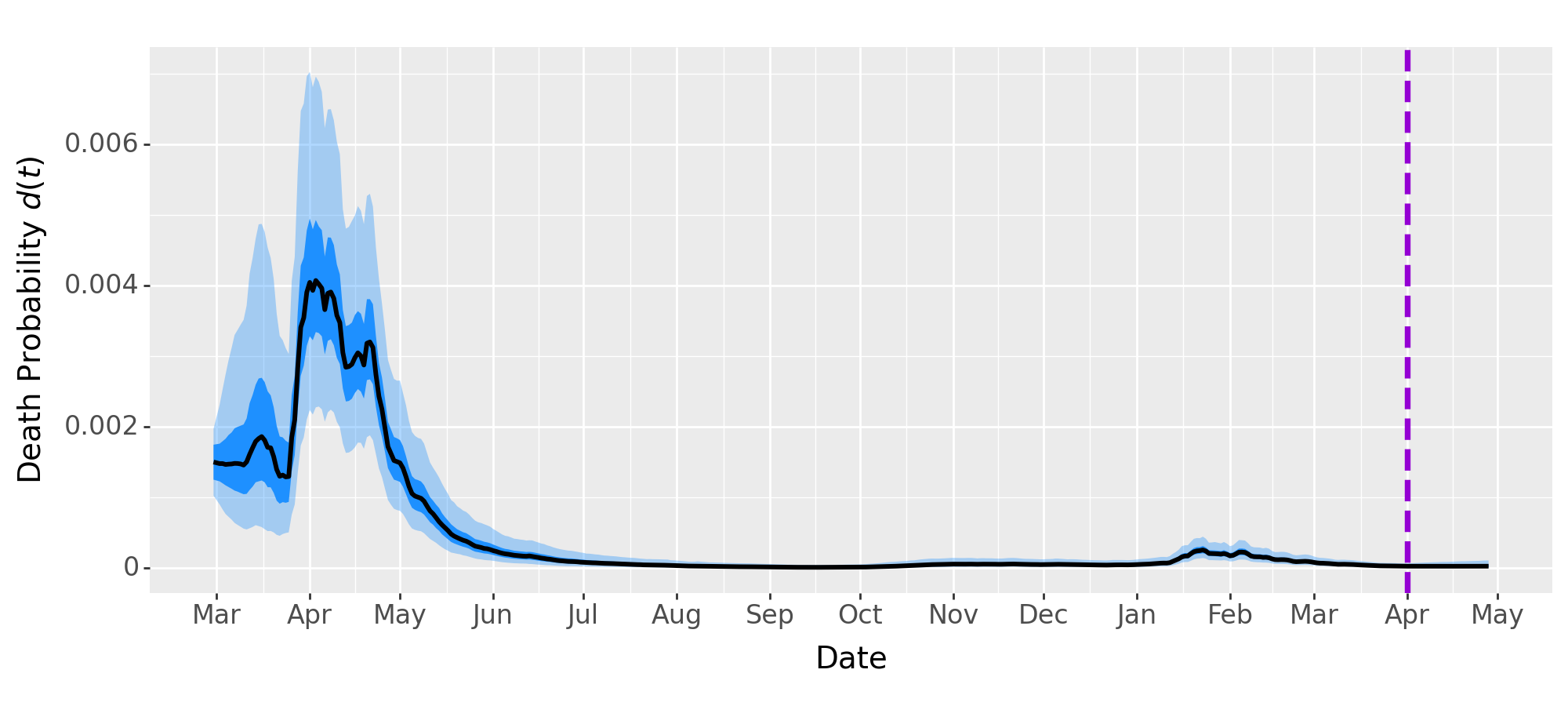}
  \end{center}
	\caption{\footnotesize {\bf Model fitting with COVID-19 data in Ireland}. The top row shows the estimation of the daily number of new cases. The middle row shows the cumulative deaths.  The bottom row shows the time-varying death probability. The observed data used in the KSF are represented by the orange dot, while the red dots are the observed data not used. The solid black line, the dark blue and light blue area correspond to the posterior median,  the 50\% and 95\% credible intervals, respectively. The vertical dashed dark violet  line corresponds to the last date of the
observations, after that the model is run 28 days forward.}
\label{Fig5}
\end{figure}

Fig~\ref{Fig5} illustrates the model fit with the daily number of reported cases. Two small peaks are noticeable: one in the middle of April 2020, and another in mid-November 2020. Additionally, there is a substantial wave in early January 2021. The model demonstrates an impressive fit to the data, accurately capturing the three waves, with observations falling within the 95\% credible interval on most days. The only exception occurs around the middle of January 2021, when there is a sudden reduction in the number of cases observed. This can be attributed to the fact that Gaussian noise results in gentle changes, making it more difficult to detect a potential change point. Beyond March 31st, 2021, the model is projected forward for 28 days (4 weeks) predictions; data from April 1st, 2021 (depicted by the red dot), were not included in the model. As we can observe in Fig~\ref{Fig5}, the actual data from April 1st, 2021, fall within the range of the 50\% credible intervals of the forecast region. 

The bottom plot of  Fig~\ref{Fig5} shows an increase in death probability (median) from over 0.15\% in March to 0.4\% early in April 2020 when there were only school closures. After the intensive lockdown measures imposed by the government from April 2020 to June 2020, the death probability drops below 0.01\%. An interesting fact is that even though we observe a high increase in the number of cases during January 2021, it does not lead to a significant increase in the death probability, suggesting a considerable degree of improvement in clinical outcomes.

\subsubsection{Effective reproduction number}

The effective reproduction number ($R_{\mathrm{eff}}(t)$) is a crucial parameter in epidemiological models that plays a key role in understanding and predicting the spread of infectious diseases. It refers to the average number of secondary infections caused by a single infected individual during their infectious period in a population where some individuals are immune or other interventions have been implemented. In general, when $R_{\mathrm{eff}}(t) > 1$, the number of observed cases will increase, while it will decrease if $R_{\mathrm{eff}}(t) < 1$. A value of $R_{\mathrm{eff}}(t) = 1$ signifies that the disease is endemic. Consequently, policymakers can determine whether to relax or strengthen control measures based on whether $R_{\mathrm{eff}}$ falls below the self-sustaining threshold of $1$.  Using the method of the next generation matrix~\cite{van2002reproduction}, we can express $R_{\mathrm{eff}}$  as:
\begin{adjustwidth}{-0.5in}{0in} 
\begin{align}
R_{\mathrm{eff}}(t) &= \beta_{t} \bigg[ (f_{as} - 1)((r_{si} - 1)f_{si}(\tau_c - \tau_l) + (r_{pi} - 1)f_{st}(\tau_c - \tau_l + \tau_r)) \notag \\
&\quad + \tau_d(f_{as} (r_{as} - r_{si}f_{si} - r_{pi}f_{st}+ f_{si} + f_{st} - 1) + (r_{si} - 1)f_{si} + (r_{pi} - 1)f_{st} + 1)\bigg]\times\dfrac{S_{t}}{N}.
\end{align}
\end{adjustwidth}

\begin{figure}[!ht]
	\begin{center}
            \includegraphics[scale=0.47]{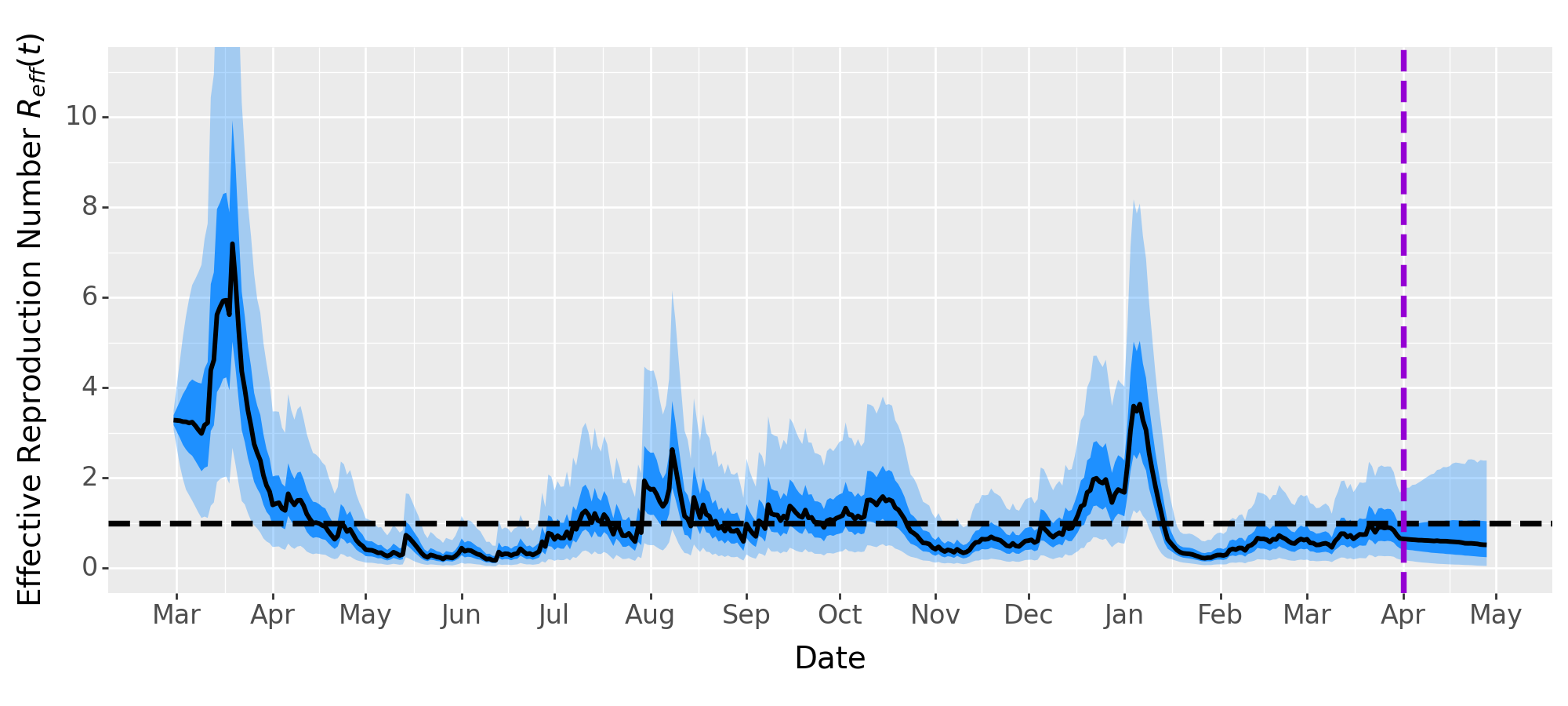}
            \includegraphics[scale=0.47]{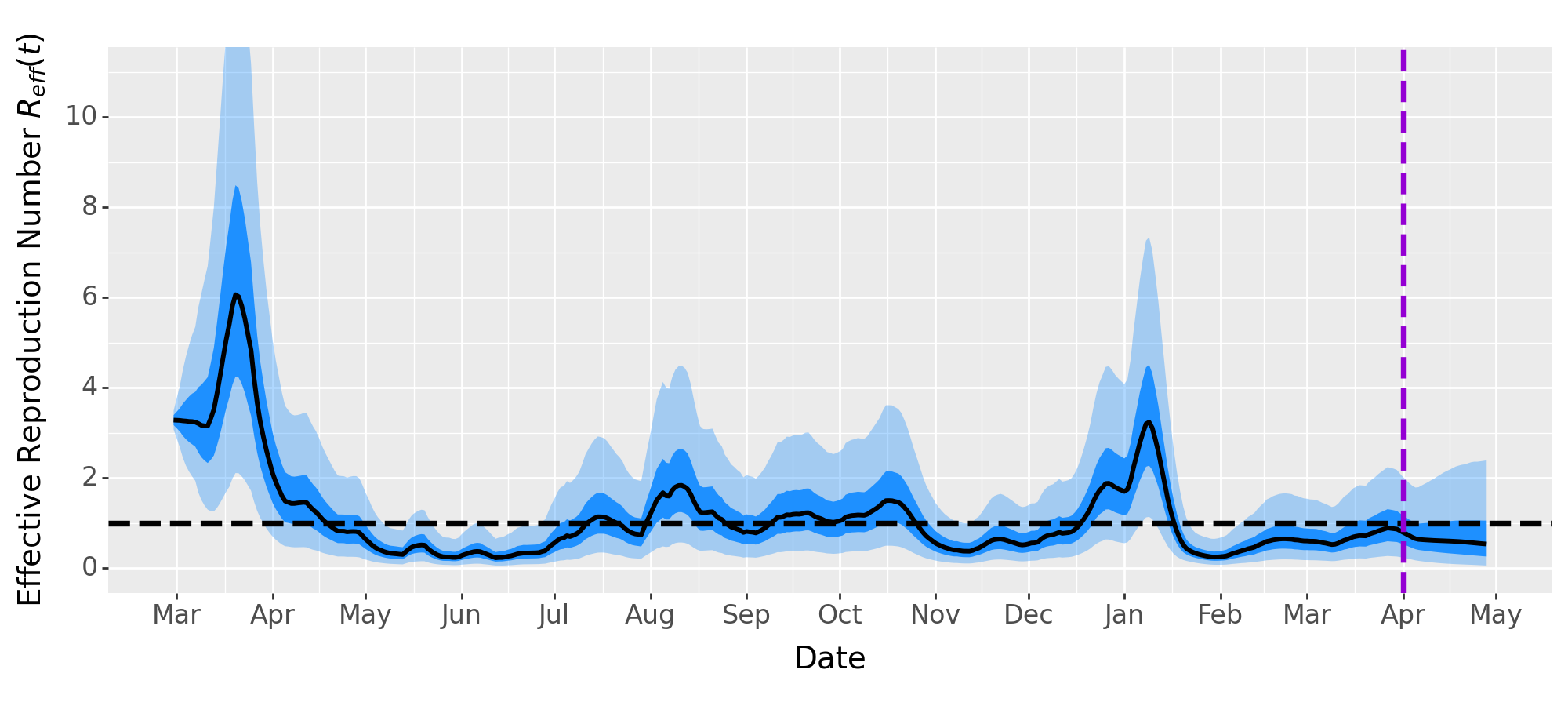}
  \end{center}
	\caption{\footnotesize {\bf Filtered estimates of the effective reproduction number}.  The top graph shows the daily estimates of $R_{\mathrm{eff}}(t)$ on a daily basis, while the bottom graph presents the estimates averaged over a weekly period. The black dashed lines are the value of $R_{eff}(t)=1$. The solid black line,  the dark blue and light blue area correspond to the posterior median, the 50\% and 95\% credible intervals, respectively. The
estimates after the last dashed dark violet line are 28-day predictions.}
\label{Fig6}
\end{figure} 

Fig~\ref{Fig6} illustrates the daily effective reproduction number ($R_{\mathrm{eff}}$) along with its weekly average. As expected, fluctuations in the epidemic are reflected in this measure. The 7-day rolling average of $R_{\mathrm{eff}}(t)$ provides a clearer depiction of the overall trend in disease spread and speed, compared to daily values alone, which may be influenced by factors like reporting delays or weekend effects. Initially, our model estimates $R_{\mathrm{eff}}(t)$ to fall within the range of $3.06$ to $3.5$, consistent with findings from prior studies~\cite{jaouimaa2021age, cazelles2021dynamics}.  It's observed that there is a notably high value of $R_{\mathrm{eff}}$ during the early stages of the epidemic, indicating possible underreporting or undetected cases due to limited testing capacity leading to an overestimation of $R_{\mathrm{eff}}(t)$. Furthermore, the impact of government lockdown measures and vaccination efforts is directly observable in the trend of $R_{\mathrm{eff}}(t)$. Predictions regarding the estimate of $R_{\mathrm{eff}}(t)$ exhibit high variability over time, with a median value slightly below 1, in accordance with the modest reduction in the number of reported cases observed in a Fig~\ref{Fig5}.

\begin{figure}[!ht]
	\begin{center}
            \includegraphics[scale=0.476]{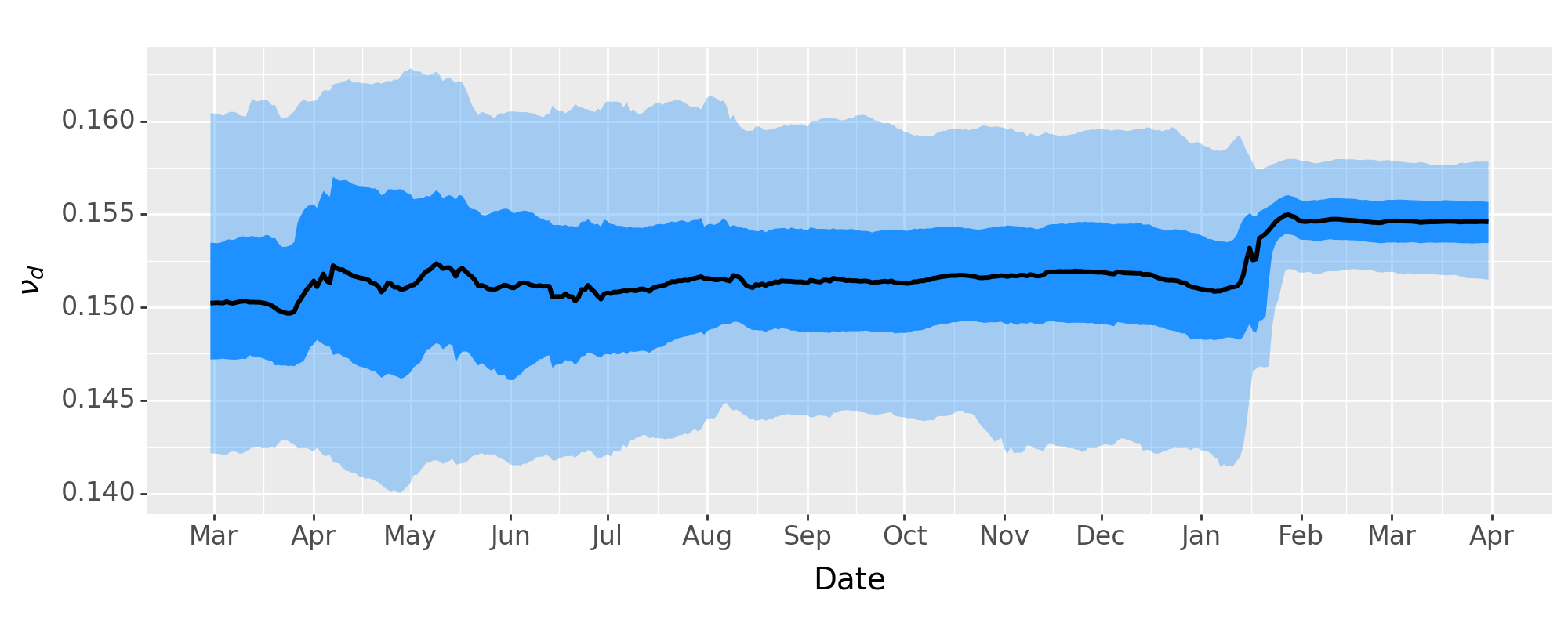}
            
            \includegraphics[scale=0.476]{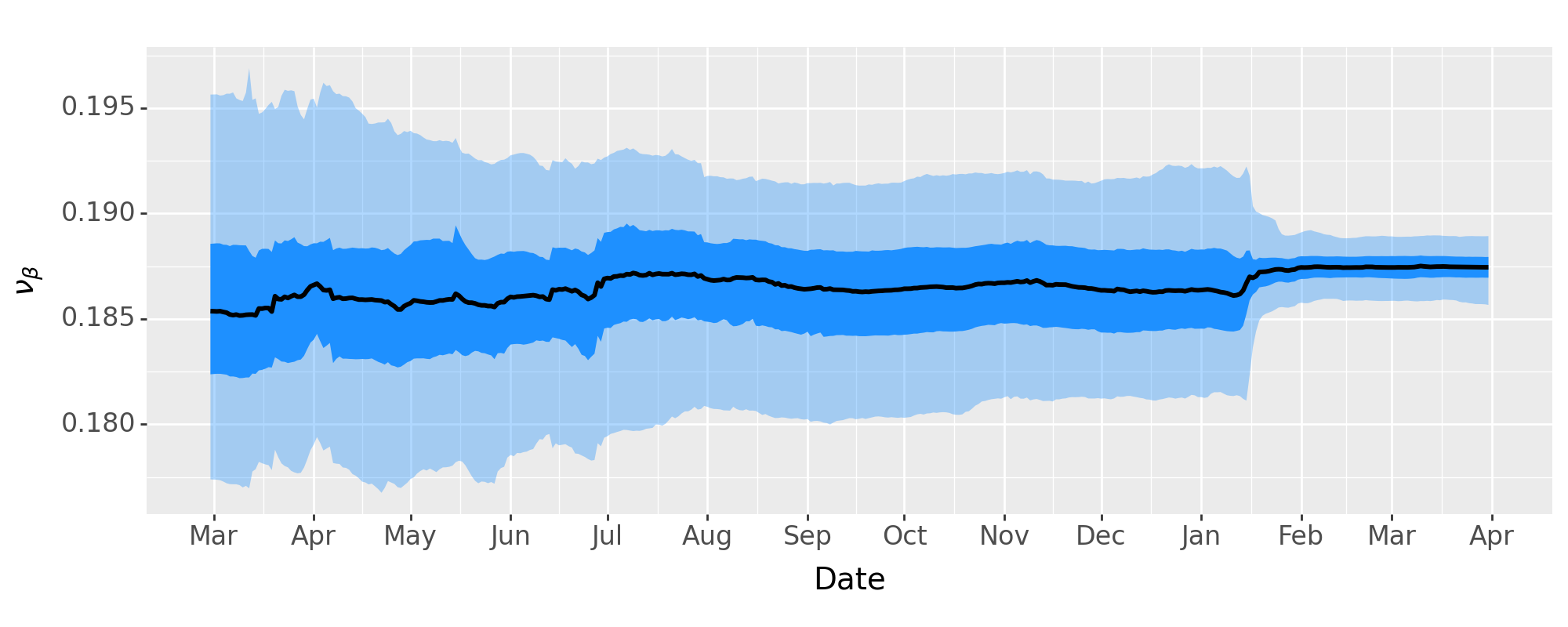}
            
            \includegraphics[scale=0.476]{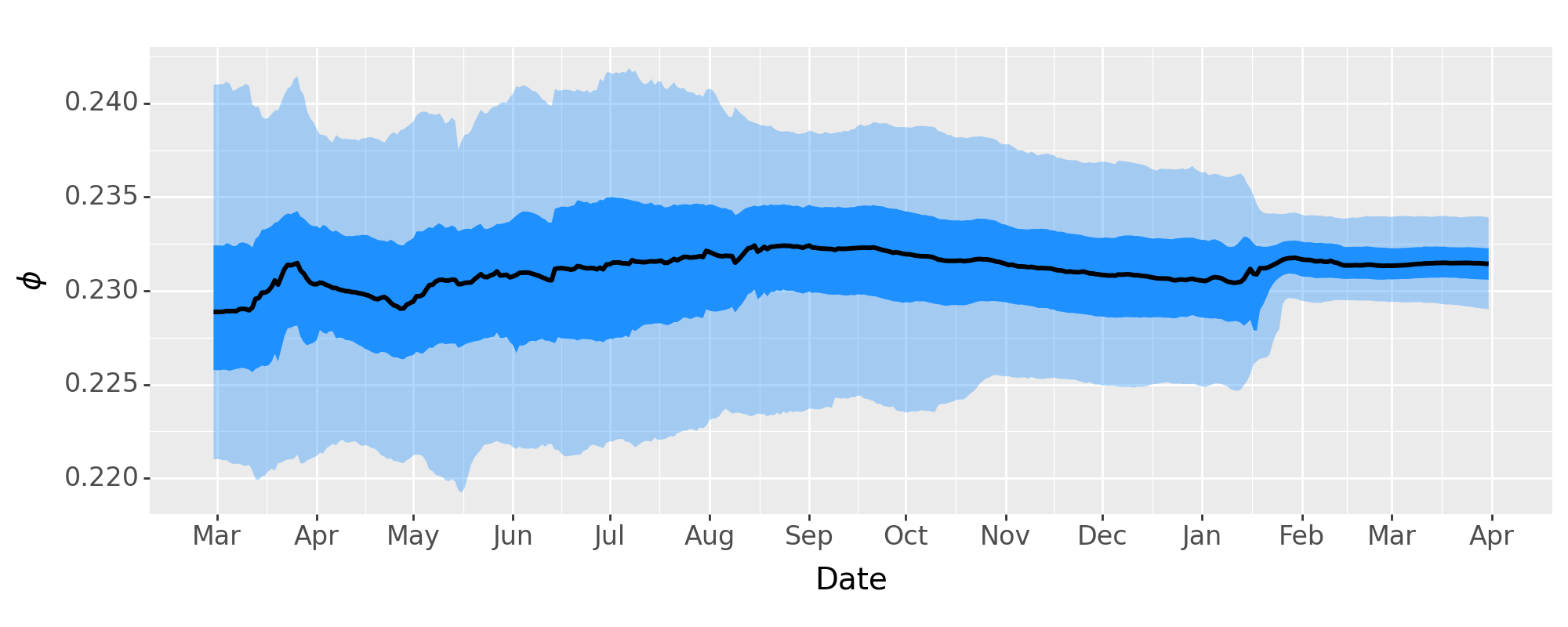}
  \end{center}
	\caption{\footnotesize {\bf Filtered estimate of the parameter using KSF algorithm}. The plots show the estimates for  $\nu_{d}$  (top), $\nu_{\beta}$  (middle) and $\phi$ (bottom).   The solid black line, the dark blue and light blue area correspond to the posterior median,  the 50\% and 95\% credible intervals, respectively.}
 \label{Fig7par_cov_ie}
\end{figure}

The KFS algorithm was also used to make inferences on the hyperparameters of the model, as shown in Fig~\ref{Fig7par_cov_ie}. It can be observed that the estimation tends to improve as more data become available. However, in mid-January 2021, a drastic reduction in the number of observed cases introduced a shock into the system. This led to the parameters deviating from their prior density, causing a depletion of the particle set in the filter. Despite this, the issue was mitigated in our case due to the large number of particles used.

\begin{table}[!ht]\label{modelparpost}
\centering
\caption{\footnotesize{\bf Prior and Posterior distribution of the hyperparameter at $T=396$ days}. Note that the prior of   $\nu_{d}$  and  $\nu_{\beta}$  are the square value but the posterior is the estimate of the parameters themselves.} \label{Table2}
\begin{tabular}{c| l| l}
\hline
\textbf{Parameter} &\textbf{Prior}&\textbf{Posterior median (95\% CI)}\\
\hline\hline
$\nu^2_d$ &  $\mathcal{IG}(80, 0.02)$ &  0.154 (0.141, 0.158)\\
$\nu^2_\beta$ &  $\mathcal{IG}(70, 0.02)$ & 0.187 (0.186, 0.189)\\
$\phi$ & $\mathcal{IG}(35, 0.2)$ & 0.231 (0.230, 0.234)\\
\hline

\end{tabular}
\end{table}

\section{Conclusion}\label{sec5}
In this paper, we have comprehensively reviewed Bayesian inference using Sequential Monte Carlo and its application in real-time disease modeling, highlighting their significant advantages over traditional deterministic models and MCMC approaches. SMC methods, with their capacity for online inference and adaptation to evolving disease dynamics, offer a robust framework for tracking and predicting the spread of infectious diseases. The integration of kernel density approximation techniques within stochastic SEIR models has proven effective in monitoring time-varying and static parameters of the disease spread dynamic, thereby providing valuable insights for timely public health interventions. Our examination of case studies, including simulations with synthetic data and analysis of real-world COVID-19 data from Ireland, underscores the practical benefits of SMC methods the instantaneous reproduction number of the infection. This method enables continuous updates of the estimation upon arrival of new observations, which is essential for responding to the fast-changing landscape of disease outbreaks. By facilitating real-time adaptation to new information, SMC approaches enhance the accuracy and responsiveness of epidemic modeling, ultimately supporting more effective public health strategies. However, epidemiological modeling based on compartmental models like SEIR can suffer from poor estimation at the early stage of disease transmission as data may not be informative. To improve early-stage modeling, future efforts will focus on refining our approach by integrating diverse modeling techniques to enhance prediction accuracy and handle uncertainties more effectively.

\section*{Acknowledgements}
Dhorasso Temfack and Jason Wyse's work was supported through 3EX, a Science Foundation Ireland Frontiers for the Future Project 21/FFP-P/10123.

\nolinenumbers

%
%
%






\end{document}